\documentclass[sigconf]{acmart}
\usepackage[utf8]{inputenc}
\usepackage{graphicx}
\usepackage{subcaption}
\usepackage{CJKutf8}
\usepackage{balance}
\usepackage[table]{xcolor}

\AtBeginDocument{%
  }

\copyrightyear{2026}
\acmYear{2026}
\setcopyright{rightsretained}
\acmISBN{xxxx-xxxx}
\begin{document}

\title[The Power of AI in Qualitative Research: Semi-Structured Interviews]{Harnessing the Power of AI in Qualitative Research: Role Assignment, Engagement, and User Perceptions of AI-Generated Follow-Up Questions in Semi-Structured Interviews}
\author{He Zhang}
\orcid{0000-0002-8169-1653}
\affiliation{%
  \institution{College of Information Sciences and Technology, The Pennsylvania State University}
  \city{University Park}
  \state{Pennsylvania}
  \country{USA}
  \postcode{16802}
}
\email{hpz5211@psu.edu}

\author{Yueyan Liu}
\authornote{Both authors contributed equally to this reserach.}
\orcid{0009-0002-8591-1374}
\affiliation{%
  \institution{Tsinghua University}
  \city{Beijing}
  \country{China}
  \postcode{100084}
}
\email{yueyan_liu@mail.tsinghua.edu.cn}

\author{Xin Guan}
\authornotemark[1]
\orcid{0009-0001-6055-1555}
\affiliation{%
  \institution{Columbia University}
  \city{New York}
  \state{NY}
  \country{USA}
  \postcode{}
}
\email{xg2413@tc.columbia.edu}

\author{Jie Cai}
\authornote{Corresponding author.}
\orcid{0000-0002-0582-555X}
\affiliation{%
  \institution{Nankai University}
  \city{Tianjin}
  \country{China}
  \postcode{300350}}
\email{jie-cai@mail.tsinghua.edu.cn}

\author{John M. Carroll}
\orcid{0000-0001-5189-337X}
\affiliation{%
  \institution{College of Information Sciences and Technology, The Pennsylvania State University}
  \city{University Park}
  \country{USA}
  \postcode{16802}}
\email{jmc56@psu.edu}


\begin{abstract}

Semi-structured interviews highly rely on the quality of follow-up questions, yet interviewers' knowledge and skills may limit their depth and potentially affect outcomes. While many studies have shown the usefulness of large language models (LLMs) for qualitative analysis, their possibility in the data collection process remains underexplored. We adopt an AI-driven ``Wizard-of-Oz'' setup to investigate how real-time LLM support in generating follow-up questions shapes semi-structured interviews. Through a study with 17 participants, we examine the value of LLM-generated follow-up questions, the evolving division of roles, relationships, collaborative behaviors, and responsibilities between interviewers and AI. Our findings (1) provide empirical evidence of the strengths and limitations of AI-generated follow-up questions (AGQs); (2) introduce a Human-AI collaboration framework in this interview context; and (3) propose human-centered design guidelines for AI-assisted interviewing. We position LLMs as complements, not replacements, to human judgment, and highlight pathways for integrating AI into qualitative data collection.
\end{abstract}


\begin{teaserfigure}
    \centering
    \includegraphics[width=0.9\linewidth]{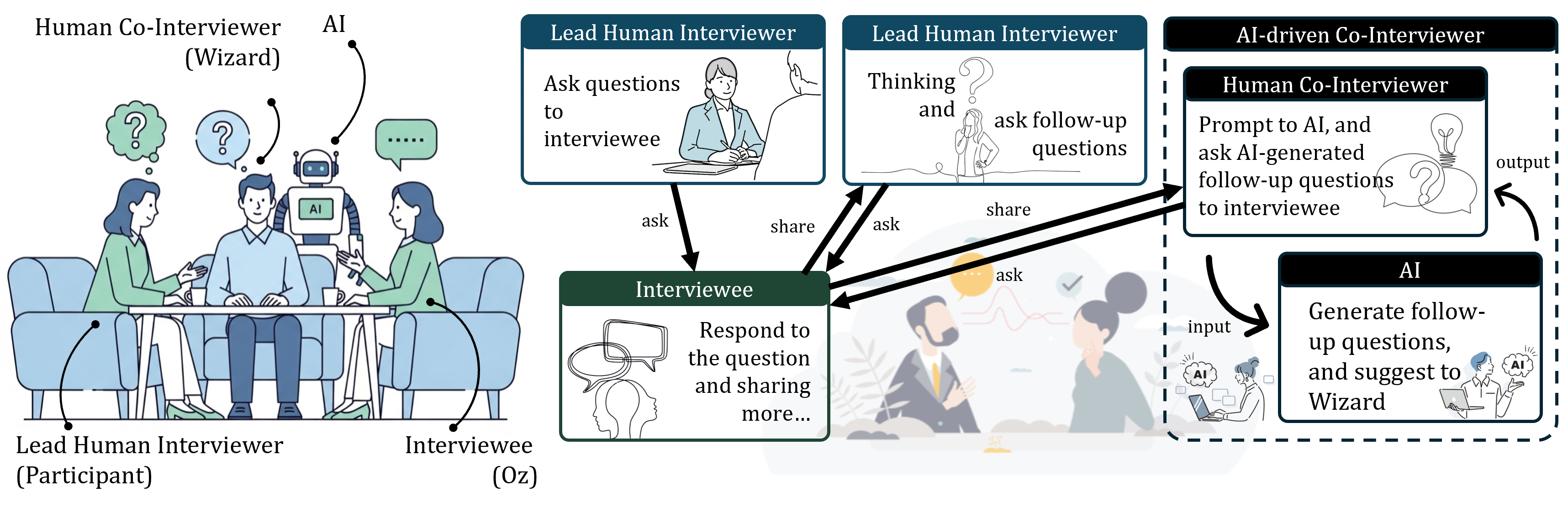}
    \caption{Schematic of the AI-driven Wizard-of-Oz method used in this study. The left panel shows the role configuration in the interview setup, where the Human co-interviewer's actual questions are generated by the AI. The right panel illustrates the interaction flow among the different parties during the interview.}
    %
    \label{fig:teaser}
\end{teaserfigure}

\maketitle

\section{Introduction}
\label{sec:intro}

Semi-structured interview is a widely used method in qualitative research~\cite{roulston2010}, uniquely positioned between structured questionnaires and entirely open-ended discussions. This approach combines predefined guiding questions with the flexibility for interviewers to explore emergent topics based on participants' responses~\cite{flick2006,charmaz2006}. Consequently, semi-structured interviews enable researchers to deeply investigate participants' experiences, attitudes, and needs, offering rich narrative insights that structured methods typically miss~\cite{adams2015}. This emphasis on participant-driven narratives facilitates unexpected discoveries and theoretical advancements~\cite{gero2025creativewriters, ballou2023needfrustration}.

Recently, the advent of artificial intelligence, especially large language models (LLMs), has introduced significant opportunities and challenges to qualitative research practices~\cite{10.1145/3706598.3713120,10.1145/3711000}. LLMs, capable of generating coherent and contextually relevant text, have already been extensively utilized in programming assistance~\cite{10.1145/3613904.3642773,10.1145/3696630.3728603,10.1145/3706598.3714002,10.1145/3706598.3714154} and automated summarization tasks~\cite{10.1145/3626772.3661346,10.1145/3706599.3720155,10669798}. In qualitative data analysis specifically, researchers have begun exploring LLMs for coding, thematic extraction, and synthetic data generation~\cite{10.1145/3613904.3641960,hayes2025conversing,zhang2023qualigptgpteasytousetool,10.1145/3613904.3642002,ZHANG2025100144}. Existing studies demonstrate that LLMs can effectively automate open coding and thematic analyses, substantially improving efficiency and consistency when handling extensive qualitative datasets~\cite{zhang2025exploring,parfenova-etal-2025-text,10.1145/3617362}. This automation potentially mitigates traditional limitations in qualitative analysis, such as high subjectivity, time-intensity, and limited scalability. As the performance of AI tools continually improves, researchers' attitudes have gradually shifted from skepticism to cautious acceptance, particularly within the Human-Computer Interaction (HCI) community~\cite{10.1145/3706598.3713726}, provided ethical standards and researcher oversight are maintained. Nonetheless, this acceptance remains contested. A substantial body of qualitative scholarship maintains that interviewing and interpretation are inherently human, relational practices that should not be delegated to automated systems, raising epistemological and ethical objections to AI involvement in qualitative inquiry~\cite{jowsey2025we,teixeira2025generative}. Rather than positioning our work within this binary, of wholesale adoption versus outright rejection, we argue that the more productive question is not whether AI belongs in qualitative research, but where, under what conditions, in what role, and under whose authority it can be responsibly bounded~\cite{greenhalgh2026reflexive,wise2026ai}.

Furthermore, despite these promising developments in the qualitative analysis stages, limited attention has been paid to integrating AI directly into the data collection phase, particularly during the interview process. A recent literature review highlights a noticeable gap, emphasizing that although researchers have extensively studied AI's role in data analysis, few studies have explored how AI can provide real-time support during interviews, with most focusing instead on demonstrating the practical capabilities of LLMs~\cite{geiecke2024conversations,10.1145/3701188} or as an awesome conversational AI chatbot in a particular case~\cite{10.1145/3652988.3673932,10.1145/3706598.3714196}. 

Semi-structured interviews are complex activities, not merely instruments for extracting pre-existing information. They are interactional settings in which the interviewer and interviewee jointly produce meaning through dialogue, clarification, and interpretation \cite{dejonckheere_semistructured_2019}. Conducting such interviews, therefore, requires more than following an interview guide. It involves building rapport with participants \cite{rowland_using_2019}, asking responsive follow-up questions \cite{robinson_probing_2023}, and exercising ongoing ethical judgment as sensitive issues and relational dynamics emerge during the conversation \cite{brinkmann_confronting_2005, hewitt_ethical_2007}. While most factors are contextual and hard to be supported by LLMs, asking follow-up questions based on interviewees' responses and a designed interview protocol might show an opportunity as they are highly language-based. We therefore examine this narrowly scoped, language-based sub-task, follow-up question generation, as a site where AI may assist while humans retain interpretive and ethical authority, treating human expertise and AI support as complementary rather than substitutive.
Given that semi-structured interviews inherently rely heavily on interviewers' active listening and rapid analytical capabilities, the ability to formulate timely and insightful follow-up questions is crucial. Nonetheless, interviewers' experience and fluctuating cognitive loads often hinder effective questioning, potentially leaving critical exploratory avenues unexplored.

Addressing this gap, our study investigates the feasibility and effectiveness of integrating AI-generated follow-up questions (AGQs) into the real-time dynamics of semi-structured interviews. Unlike traditional Wizard-of-Oz approaches, where human researchers simulate AI behavior, we employ an AI-driven/Reverse Wizard-of-Oz methodology, wherein real-time AI-generated questions are introduced by a researcher acting as a co-interviewer. This reversal emphasizes AI's supportive role in enhancing interview processes. We conceptualize this configuration as a `backstage AI' model, in which the generative agent operates invisibly to the participant, serving as a hidden cognitive engine, while the human interviewer serves as the sole frontstage interface.

Specifically, we explore two central research questions:
\begin{itemize}
    \item[\textbf{RQ1:}] \textbf{\textit{How do human interviewers perceive the quality and utility of follow-up questions generated by the backstage AI?}}
    
    This question assesses whether AI-generated prompts effectively deepen the exploration of participant narratives and viewpoints. It specifically evaluates the relevance, timing, and strategic value of the AGQs from the interviewer's professional perspective.

    \item[\textbf{RQ2:}] \textbf{\textit{How does the collaboration with a backstage AI co-pilot impact the human interviewer's perceived role, agency, and interaction dynamics?}} 
    
    This question explores the shifting human-AI relationships within the interview workflow. It focuses on critical HCI dimensions, including interviewer trust, sense of control, and role delineation (e.g., shifting from question generator to evaluator) when working alongside a hidden generative agent.
\end{itemize}

By addressing these research questions, our study contributes in four significant ways: First, through immersive AI-driven/Reverse Wizard-of-Oz experiments, we empirically validate the practical value of AGQs by assessing their impact on interview depth, fluidity, participant experience, and acceptability. Second, we propose a structured framework for role allocation in Human-AI collaborative interview, identifying best practices for integrating AI while delineating clear boundaries to enhance rather than replace human interviewer expertise. Third, we provide actionable guidelines for designing AI-assisted semi-structured interview tools, emphasizing ethical considerations, privacy protection, and maintaining participant trust. Finally, we discuss the collaborative dynamics between interviewers, AI, and interviewees, analyzing how AI interventions alter interaction patterns and suggesting strategies for effective Human-AI collaboration.

\section{Related Work}
\subsection{Foundations of Semi-Structured Interviews in HCI}
Semi-structured interviews as a qualitative research method in a main method to understand users' experience and habitual patterns to facilitate system design; however, it also requires a set of skills to gain deep insights about experience and patterns. Asking appropriate follow-up questions, along with developing the interview guide, are important steps.  In the following sections, we will introduce the value of semi-structured interviews to the HCI community and the importance of follow-up questions.

\subsubsection{Theoretical Foundations and Value in HCI}

Semi-structured interviews are among the most widely used data collection methods, combining a structured approach with the flexibility of open-ended, unstructured exploration~\cite{roulston2010}. Semi-structured interviews are generally based on a predetermined framework and oriented towards a central topic that shapes the overall flow of the conversation~\cite{Magaldi2020}. The flexibility allows researchers to improvise follow-up questions based on interviewees' responses, eliciting a detailed exploration of their experiences, perceptions, and needs~\cite{adams2015}. Unlike structured surveys, semi-structured interviews encourage participants to articulate their stories and perspectives in their own words~\cite{flick2006, marshall2014}. The focus on interviewees' viewpoints provides unanticipated insights and supports exploratory analysis~\cite{charmaz2006}. In HCI research, semi-structured interviews can help identify current user practices and evaluate the impact of new technologies, thus supporting the broader goal of addressing user needs through technological solutions~\cite{blandford2013}.

Semi-structured interviews can take different forms, including one-to-one interviews, co-interviews, group interviews, and focus group interviews~\cite{Bechhofer1984}. In particular, the co-interview approach, or `the tandem interview'~\cite{Kincaid1957a}, refers to the practice in which two or more interviewers interview a single interviewee together. It has recently gained some attention, although it remains marginal compared to the traditional single-interview approach~\cite{Monforte2021}, or the broader exploration of interviewer-interviewee relationships~\cite{Verjee2024}. Co-interviews offer both benefits and challenges \cite{velardo2021}. Compared to interviews conducted by a single interviewer, this method brings diverse perspectives to the discussion and enables each interviewer to build on aspects that the other may miss, leveraging their complementary expertise and positions~\cite{rosenblatt2012,Kincaid1957a}. It can also help promote social support, opportunities for reciprocal learning, and the development of closer relationships between the researcher and the participant~\cite{velardo2021}. Additionally, co-interview allows interviewers to pursue follow-up questions on unexpected issues more effectively during the conversation~\cite{Kincaid1957b}. However, co-interview can reshape power dynamics, potentially creating imbalances within the interviewers~\cite{redmanmaclaren2014}. The logistical demands of preparation and coordination can also add complexity to the interview~\cite{velardo2021}.

\subsubsection{Implementation Practices and Interviewer Dynamics}

Conducting semi-structured interviews requires a set of skills that balance structured preparation with the flexibility to adapt during the conversation. Effective preparation requires interviewers to draw on prior knowledge of the research field to construct an interview guide that ensures core topics are adequately addressed~\cite{wengraf2001, kelly2010, kusk2025consent}. Rather than prescribing rigid question items, an interview guide should outline the key themes to be examined~\cite{arthur2003}. This structure allows for adaptability as interviewers can improvise follow-up questions in response to interviewees' answers~\cite{marshall2014,charmaz2006}. Interviewers can encourage interviewees to elaborate on their earlier points by repeating or expressing their interests~\cite{whiting2008semi}, and use specific examples~\cite{dearnley2005reflection} based on the interviewees' statements. Developing spontaneous follow-up questions is a critical element in semi-structured interviews~\cite{turner2010qualitative}. This follow-up strategy can not only enrich the depth and details of responses \cite{whiting2008semi} but also help keep the natural flow of the conversations~\cite{roulston2003} and uncover subtle nuances~\cite{rubin2005qualitative}. 

In HCI research, follow-up questions have proven especially valuable for sustaining deeper inquiry and moving from descriptive accounts to higher-level conceptual themes. This probing enriches the understanding of user experiences, providing actionable insights for HCI practice, including system design in areas, such as data privacy~\cite{martius2025out, kusk2025consent}, healthcare~\cite{motahar2025skiers, xiong2025petreunion}, and education \cite{10.1145/3706598.3714232,10.1145/3686852.3687069}. It is particularly beneficial when addressing the needs of marginalized or vulnerable groups \cite{10.1145/3706598.3713433,10.1145/3706598.3714210}. Beyond design implications, the follow-up questions also support methodological contributions by linking emerging concepts for the development of theory and models in HCI studies~\cite{gero2025creativewriters, ballou2023needfrustration}.

However, improvising follow-up questions in practice can be challenging, such as processing information quickly when deciding how deeply to elicit \cite{price2002}, sustaining interviewee engagement and managing silence during dynamic conversations~\cite{pope1974experienced}, over-directing the interaction or speaking excessively~\cite{MYERS20072}. These difficulties become more pronounced when interviews involve sensitive topics, which require greater ethical awareness and experience~\cite{whiting2008semi}. In the HCI community, these challenges can also affect the validity of user analysis and the design of systems within their intended environments, particularly through the influence of confirmation bias and other cognitive biases~\cite{10.1145/1822090.1822094}.

With the advancement of AI, human-AI collaboration has become a growing area of interest in HCI, including its potential to enhance qualitative research methods. Building on the importance of follow-up questions in semi-structured interviews, this study investigates how LLMs might assist interviewers in generating follow-up questions during interviews, positioning this process as a form of human-AI collaborative practice.

\subsection{AI and Large Language Models Support for Qualitative Research}
Qualitative research is a cornerstone of HCI and the social sciences, providing in-depth insights into people's practices, needs, and values~\cite{blandford2016qualitative}. Yet, qualitative methods, especially semi-structured interviews, are often labor-intensive, requiring significant skill, experience, and attentional resources from researchers~\cite{whiting2008semi}. To address these challenges, scholars have explored technological interventions to improve efficiency, reduce cognitive load, and enhance the quality of qualitative methods~\cite{redlich2014new,paulus2013digital,klein2007use,silver2014using,davidson2016speculating,palys2012qualitative}. 

\subsubsection{Technological Supports for Interviews before LLMs}
Before the advent of generative models like LLMs, researchers experimented with various computational tools to support interview processes~\cite{jain2021survey}. The most relevant developments came from the field of natural language processing (NLP), where advances in neural architectures such as Transformers~\cite{vaswani2017attention}, BERT~\cite{devlin-etal-2019-bert}, subsequent pretrained sequence-to-sequence models like BART~\cite{lewis-etal-2020-bart} and Text-to-Text Transfer Transformer~\cite{JMLR:v21:20-074} created the conditions for automatic question generation~\cite{heilman-smith-2010-good}. Approaches included rule-based question generators~\cite{b-etal-2020-automatic,debnath2020designing} and scripted dialogue systems~\cite{10.1145/505282.505285,10.1145/3637320}, which advanced rapidly. These systems showed promise in producing standardized prompts and scaffolding semi-structured conversations, but their reliance on predefined rules or narrow training corpora often limited adaptability to diverse interview settings.

In parallel, the HCI community explored how computational systems could mediate or scaffold interviews in practice. Early work investigated embodied conversational agents and virtual interviewers in employment or survey contexts~\cite{10.1145/2493432.2493502,10.1145/3613904.3642707,10.1145/3232077}, while other studies experimented with chatbot-assisted surveys and conversational probes to elicit richer qualitative data~\cite{10.1145/3381804,10.1145/3411764.3445569,10.1145/3290605.3300705,10.1145/3411764.3445116,10.1145/3313831.3376131}. Building on both NLP innovations and HCI explorations, subsequent research began to introduce more powerful AI-driven systems into interview settings. More recent projects examined AI-powered tools~\cite{10.1145/3382507.3418839} that provide real-time follow-ups or theory-driven probes, highlighting opportunities for efficiency and consistency, but also surfacing concerns over interviewer control, rapport, and ethical responsibility~\cite{liu2023speech,10.1145/3581641.3584051,biswas2024hi}.

\subsubsection{LLM-Supported Interviews}
Recent advances in LLMs have reshaped possibilities for supporting qualitative research. Methods such as prompt engineering, structured input-output formatting~\cite{liu2024llms,ZHANG2025100144}, chain-of-thought prompting~\cite{NEURIPS2022_9d560961}, retrieval-augmented generation (RAG)~\cite{gao2024retrievalaugmentedgenerationlargelanguage,10762977}, and reinforcement learning with human feedback (RLHF)~\cite{bai2022traininghelpfulharmlessassistant} can improve reliability and contextual grounding. As LLMs can generate contextually relevant text and sustain coherent dialogue, they are well-suited for interactive settings like interviews~\cite{10.1145/3652988.3673932}. Prior studies have explored LLMs primarily in the analytic stages of qualitative research, for example, in transcription~\cite{mojadeddi2024automated}, thematic coding~\cite{christou2024thematic,10.1145/3613904.3642002,ZHANG2025100144}, and content analysis~\cite{wang2025lata,overney2024sensemate, dai2023llm}. These works highlight LLMs' capacity to accelerate analysis, improve consistency, and reduce researcher burden. However, less attention has been given to their role in data collection, particularly in semi-structured interviews. 

Some scholars have examined LLMs as conversational agents or assistants in interview-like settings, showing that they can generate follow-ups or clarifications that enrich the dialogue~\cite{10.1145/3686215.3688377,spangher-etal-2025-newsinterview, liu2025aiinterviews}. Yet some studies rarely consider how AI-generated questions affect the interview's flow, interviewer authority, or interviewees' trust, particularly from the interviewers' perspectives~\cite{spangher-etal-2025-newsinterview}. In addition, Human-AI collaboration patterns received less attention in the co-interview setting with two human interviewers involved, which complicated the collaboration in the three-party dynamic relationship of ``interviewer - interviewee - AI''~\cite{liu2025aiinterviews}. In particular, gaps remain in understanding how AI should balance initiative versus deference, how its phrasing and timing shape rapport, and how researchers can maintain control and accountability in hybrid human-AI interview scenarios.

Our study addresses these gaps by investigating the usefulness, risks, and interactional dynamics of AI-generated follow-up questions in semi-structured interviews. Specifically, we focus on when and why such questions are perceived as valuable, how they influence the lead interviewers' experience, and what design implications emerge for future human-AI collaborative tools. By shifting attention from analysis to the interview process, we contribute to ongoing HCI discussions on how to design AI supports that preserve the interpretive richness of qualitative methods while enhancing their practicality and scalability.

\section{Methods}

\subsection{Study Design}
We employed an immersive AI-driven/Reverse Wizard-of-Oz~\cite{10.1016/0950-7051(93)90017-N} methodology to investigate the effectiveness and user experience of integrating LLMs into interview processes. This method allowed us to evaluate participants' reactions to AI-generated interventions while maintaining authentic interaction dynamics. Each study session involved three explicitly defined roles: one genuine recruited participant, one researcher acting as a simulated interviewee (``Oz''), and another researcher functioning as a co-researcher (``Wizard''). The genuine participant temporarily assumed the primary interviewer role, leading interactions with ``Oz''. 

All participants (Oz) have experience with semi-structured interviews. We asked them to reflect on the topics from their prior interviews and bring their interview protocols to our experiment. Meanwhile, the researcher ``Wizard'' reviewed their interview topics and selected topics relevant to the researcher's experience as potential candidates. We clarify that interview topics were selected and matched to ensure that the ``Oz'' had relevant personal experience or sufficient contextual knowledge to respond meaningfully. In this sense, the ``Oz'' responses were not pre-scripted or artificially fabricated; rather, they emerged through an interview process that incorporated several key practical characteristics of real-world interviews.

Critically, the ``Wizard'' did not independently formulate follow-up questions; instead, all follow-up questions were generated in real-time by GPT-4o, based directly on the responses provided by ``Oz''. To thoroughly explore various AI intervention modes, we systematically varied the Wizard's follow-up questioning strategies, utilizing three distinct interaction patterns: (1) occasionally interjecting single follow-up questions amidst the participant's primary questioning; (2) periodically inserting a follow-up question after every one or several questions asked by the participant; and (3) occasionally posing two or more consecutive follow-up questions without immediate opportunity for the participant to interject. These strategies aimed to simulate different frequencies and styles of AI interventions, enabling an investigation into optimal interaction modes and their respective impacts.

To mitigate potential demand characteristics and Hawthorne~\cite{adair1984hawthorne} / Halo effects~\cite{pohl2016cognitive,nisbett1977halo}, we adopted a two-phase reflection design in which participants first evaluated the follow-up questions without knowing their AI origin, and only afterwards were informed of the AI's involvement (see Section~\ref{subsec:study-procedure}). This before–after comparison allowed us to detect whether disclosure shifted their assessments. Throughout the session, the ``Oz'' and ``Wizard'' gave no cues about the perceived quality of any question, and the interview protocol explicitly invited critical and negative feedback (e.g., asking participants to identify ill-timed or irrelevant questions) to counterbalance social-desirability bias.

The core objective of this design was to assess participants' subjective experiences, cognitive perceptions, and attitudes toward collaborating with AI-generated prompts within realistic contexts, thereby capturing crucial insights for future Human-AI collaborative interview practices.

As part of our semi-structured interview protocol, participants were invited to reflect on several core dimensions of qualitative interview. First, we explored their prior experiences and practices in conducting qualitative research (e.g., ``\textit{Can you describe your previous experience conducting qualitative interviews?}'' and ``\textit{What challenges have you encountered, and how did you address them?}'').

The protocol then shifted to participants' perceptions and evaluations of Human-AI collaboration within the interview context. Here, we asked participants to assess their experience with AI-mediated support (e.g., ``\textit{How did you feel about my [the Wizard's] involvement in your interview?}''), and to evaluate the perceived relevance, pacing, and utility of the AGQs (e.g., ``\textit{How would you evaluate the follow-up questions I asked—such as their relevance, pacing, or how thought-provoking they were?}''). We also probed whether participants attributed their experience more to the content of the follow-up questions or to their delivery and timing.

Additional topics included role boundaries and interviewer agency (e.g., ``\textit{How did my [Wizard] participation affect your sense of agency or the dynamic of interviewer-interviewee roles?}''), as well as cognitive strategies for adapting to AI interventions (e.g., ``\textit{After we introduced a follow-up question, what was your immediate thought process or next step?}'').

After disclosing the AI's role, we invited participants to reflect on whether this awareness shifted their assessment of question quality or relevance (e.g., ``\textit{Knowing these were AI-generated, does your assessment change?}''), and to consider the relative importance of question content versus human factors such as delivery, tone, and emotional nuance.

Finally, to elicit perspectives on future practices, we asked participants to articulate their preferences for human-AI collaboration in generating follow-up questions (e.g., ``\textit{Would you prefer to collaborate with a human or with an AI in future interviews?}''), and to share their reasoning regarding the desired balance between human and AI contributions in qualitative interview.

\subsection{Study Procedure}\label{subsec:study-procedure}
\begin{figure*}[htbp]
    \centering
    \includegraphics[width=0.86\linewidth]{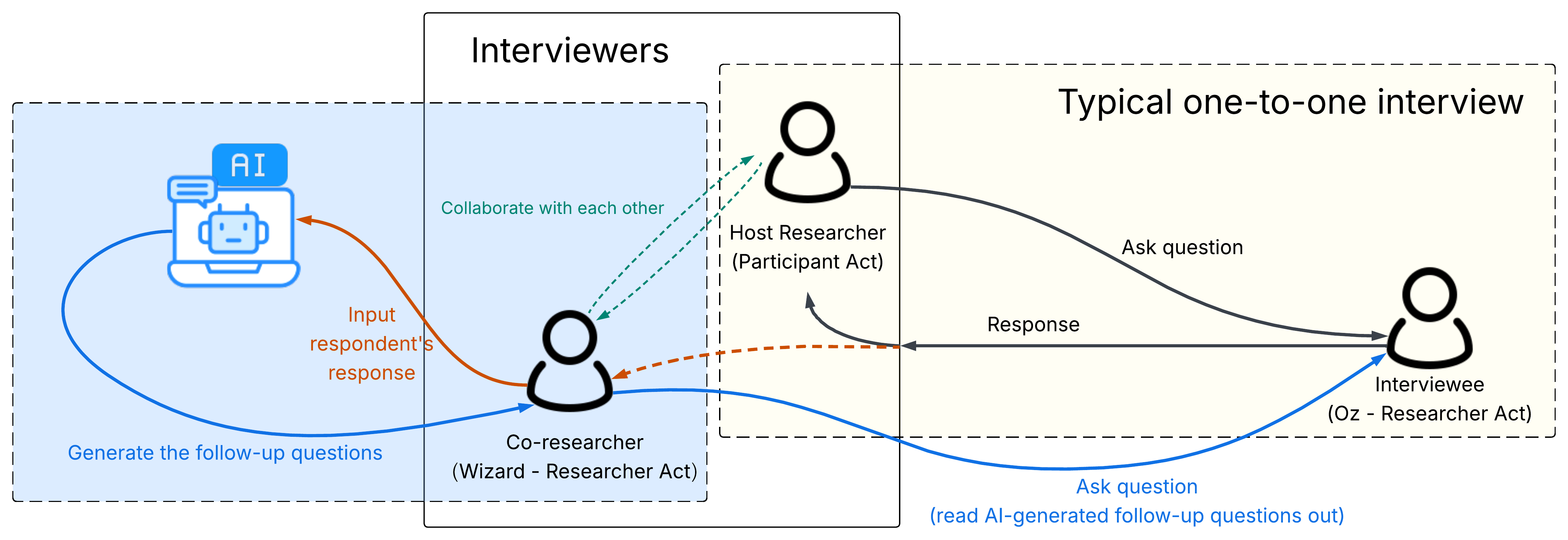}
    \caption{Study Procedure. The figure illustrates the AI-driven/Reverse Wizard-of-Oz study design. On the right, a typical one-to-one semi-structured interview is shown, where the host researcher interacts with the interviewee. On the left, the co-researcher secretly collaborates with the AI system: the co-researcher inputs the interviewee's responses into the AI, receives AGQs, and selectively voices them to the interviewee.}
    \label{fig:studyProcedure}
\end{figure*}
\vspace{-0.2cm}
Participants were asked to provide interview topics and outlines based on themes familiar to or of interest to them prior to the session. Each session began with a brief orientation designed to enhance participant immersion. Participants were informed explicitly about the session's objectives, structure, ethical considerations, recording procedures, confidentiality assurances, the right to withdraw at any time, and compensation details.

The session commenced with an ice-breaking activity, during which participants briefly shared their background, experiences in qualitative research, previous interview strategies, and challenges encountered in past interview situations.

The main experimental session involved a 15--30 minute semi-structured interview where the genuine participant, acting as the interviewer, interacted with ``Oz'' under the supervision of the ``Wizard.'' The ``Wizard'' was tasked with managing the GPT-4o workflow, which was pre-configured with a persona specializing in qualitative inquiry. Crucially, to maintain experimental consistency, we utilized a straightforward, template-based prompt rather than complex chain-of-thought engineering. The Wizard consistently input the following command: \textit{`Please give me three follow-up questions based on responses: [participant's responses],'} thereby instructing the model to generate relevant probes based solely on the immediate dialogue context.

During the interview, the ``Wizard'' strategically inserted GPT-generated follow-up questions following the three distinct interaction patterns described in the study design, ensuring interventions were contextually coherent and naturally integrated.

After completing the semi-structured interview, a reflective discussion took place without initially disclosing GPT-4o's involvement. Participants evaluated the ``Wizard's'' interventions, specifically assessing the relevance, timing, and collaborative dynamics of the follow-up questions. Participants also reflected on how the co-researcher's presence impacted their control over the interview, cognitive load, and collaborative experience.

Subsequently, participants were informed that all follow-up questions were generated by GPT-4o, prompting a secondary reflection phase. Here, participants reassessed their perceptions of AI-generated prompts, particularly discussing whether the quality and presentation of the questions or human interaction factors (such as tone and pacing) influenced their experiences more significantly. Further discussions explored participants' preferences regarding future Human-AI collaborative scenarios, anticipated challenges with AI-assisted interview, and suggestions for refining GPT-assisted qualitative interviews.

Additionally, in select sessions, GPT-generated follow-up questions were visually presented to participants to investigate cognitive load and practical usability implications. Throughout the experimental process, we closely replicated real-world qualitative interview workflows to maintain ecological validity.


\subsection{Data Collection \& Analysis}
All study sessions were conducted remotely using secure video conferencing platforms (e.g., Zoom). With informed consent, both audio and video of each interview were recorded for research purposes. Recordings were subsequently transcribed verbatim and anonymized before analysis; all personal identifiers were removed or replaced with pseudonyms to ensure confidentiality.

Each session was recorded in its entirety, capturing both participant and AI interactions. Our analysis followed a reflexive thematic analysis approach~\cite{Braun01012006,braun2021thematic}, which treats themes not as units to be reliably identified, but as analytic constructs developed through the researchers' active, interpretive engagement with the data. This orientation fits our study's interpretive aim of understanding how participants made sense of AI-generated follow-up questions, particularly the shifts in their reflections before and after AI disclosure. Three researchers, each with prior experience in qualitative interview, independently open-coded a subset of transcripts to identify preliminary codes related to perceptions, experiences, and responses to AGQs. We specifically compared participant reflections before and after the disclosure of the AI's involvement to explore the impact of this revelation on their attitudes and feedback.

Rather than calculating inter-rater reliability (IRR) or working toward a fixed codebook, the four-member research team met regularly (one to two times per week) to discuss these codes collaboratively. Consistent with a reflexive orientation, divergences between coders were treated as analytically generative rather than as errors to be eliminated: differing readings prompted us to return to the data, refine our interpretations, and develop higher-level themes. Themes were thus constructed through interpretation and dialogue across the team, and were iteratively reviewed and revised over several weeks until they offered a coherent account of the data. To illustrate this analytic chain with one example developed more fully in our findings (Section~\ref{sec:timing}): a participant's account of the assistant jumping ``straight to the main question'' before they were ready (P13) led us to an initial code capturing the assistant's tendency to insert questions too early. Through team discussion, this was related to other participants' accounts of losing control over pacing (e.g., P11, P14), and gradually developed into a higher-level theme about the timing of AI suggestions, which in turn informed our interpretation of how delivery and timing, not content alone, shape the perceived usefulness of AGQs.

In addition to coding, the research team held debriefing discussions after each interview session. These regular reflections on participant responses, AI interactions, and emerging analytical directions fed back into our developing interpretations and supported the construction of a shared, reflexive understanding of the data across the team.

\subsection{Participants}
Participants were recruited through flyers distributed across various social media platforms. A total of 59 individuals initially expressed interest in the study. To ensure relevance and quality of participation, we established specific inclusion criteria, with a primary focus on participants' experience with qualitative research methodologies. During the recruitment phase, interested individuals were asked to submit a brief summary of a topic they were both interested in and knowledgeable about, along with a corresponding interview outline.

Eligible participants were selected on a rolling, first-come, first-served basis. 
All 17 participants had prior experience conducting qualitative interviews, consistent with our inclusion criteria. Based on the experience levels defined in Table~\ref{table:demographic}, the sample comprised 10 Novice, 5 Intermediate, and 2 Advanced interviewers; among them, the 7 Intermediate and Advanced participants had published at least one peer-reviewed paper or report based on qualitative methods.

We also collected demographic data, including education level, occupation, and professional domain, for descriptive reporting purposes; however, these factors were not considered in the selection process. We excluded individuals who had no prior interview experience, those we could not contact, those who declined after initial registration, and those whose proposed interview topics did not align with scenarios the research team could realistically simulate. All potential participants were informed that not all applicants would be selected for the final study.

Recruitment proceeded iteratively and was concluded once theoretical saturation was reached, that is, when successive interviews yielded no substantially new themes, and participant feedback began to converge. The final sample consisted of 17 participants (ages 21 to 35), including 9 females and 8 males.

\begin{table*}[htbp]
\resizebox{\textwidth}{!}{%
\begin{tabular}{c|ccccc}\hline
\begin{tabular}[c]{@{}c@{}}Participant \\ No.\end{tabular} &
\begin{tabular}[c]{@{}c@{}}Age\end{tabular} &
\begin{tabular}[c]{@{}c@{}}Gender\end{tabular} &
\begin{tabular}[c]{@{}c@{}}Education\end{tabular} &
\begin{tabular}[c]{@{}c@{}}Working Area\end{tabular} &
\begin{tabular}[c]{@{}c@{}}Experience of \\ Qualitative Method*\end{tabular} \\ \hline
P1  & 35 & Female & PhD      & Smart Home                                 & Intermediate \\
\rowcolor[HTML]{EFEFEF}
P2  & 24 & Female & Master    & Education                                  & Novice \\
P3  & 28 & Male   & Master    & Smart Home                                 & Intermediate \\
\rowcolor[HTML]{EFEFEF}
P4  & 24 & Female & Master    & Human Geography                            & Novice \\
P5  & 33 & Female & Master    & Ergonomics                                 & Novice \\
\rowcolor[HTML]{EFEFEF}
P6  & 22 & Female & Master    & Education                                  & Novice \\
P7  & 30 & Male   & Master    & Management Science and Engineering         & Novice \\
\rowcolor[HTML]{EFEFEF}
P8  & 21 & Male   & Bachelor  & Education                                  & Novice \\
P9  & 22 & Female & Bachelor  & Education                                  & Intermediate \\
\rowcolor[HTML]{EFEFEF}
P10 & 23 & Female & Master    & Social Work                                & Novice \\
P11 & 24 & Male   & Master    & HCI                                        & Advanced \\
\rowcolor[HTML]{EFEFEF}
P12 & 32 & Male   & PhD       & Usable Privacy and Security                & Intermediate \\
P13 & 21 & Female & Bachelor  & Humanities and Social Sciences             & Novice \\
\rowcolor[HTML]{EFEFEF}
P14 & 23 & Female & Master    & Psychology                                 & Novice \\
P15 & 33 & Male   & PhD       & HCI                                        & Advanced \\
\rowcolor[HTML]{EFEFEF}
P16 & 25 & Male   & Master    & Applied Linguistics                        & Novice \\
P17 & 26 & Male   & PhD       & HCI                                        & Intermediate \\ \hline
\multicolumn{6}{l}{\textbf{*Experience of Qualitative Method}}\\
\multicolumn{6}{l}{\textit{Novice}: Have participated in data collection for qualitative research, especially interviews.}\\
\multicolumn{6}{l}{\textit{Intermediate}: Have published 1-3 peer-reviewed paper or written a report based on qualitative method, especially interviews.}\\
\multicolumn{6}{l}{\textit{Advanced}: Have published more than three peer-reviewed papers based on qualitative method, especially interviews.}
\end{tabular}%
}
\caption{Participants' demographic information}
\label{table:demographic}
\end{table*}

\section{Findings}\label{sec:findings}
Our analysis presented two themes corresponding to our research questions: (1) the perceived usefulness of AGQs in semi-structured interviews, and (2) the perception of the support from the follow-up assistant in co-interview with the backstage AI. Across both, participants consistently emphasized the importance of alignment with research aims, timing sensitivity, and human control over ethical and relational dimensions. In the finding sections, the letters A(dvanced), I(ntermediate), and N(ovice) were added to the participant numbers for identification purposes.

\subsection{RQ1: Perceived usefulness of AGQs}\label{sec:value}

This section presents findings addressing RQ1, emphasizing the lead interviewers' perceived usefulness of AGQs and the factors that influence these perceptions. Overall, participants frequently highlighted the usefulness of AGQs to the interview process, improving both the content and the flow of conversation. However, perceived usefulness is influenced by factors such as contextual relevance, timing, and ethical considerations. When these conditions were met, AGQs provided scaffolding that expanded the interviewer's repertoire while maintaining human judgment; when such conditions did not hold, AI risked disrupting flow or reframing the topic, undermining the usefulness of AGQs. The complexity and dynamic usefulness of AGQs should be emphasized, as they depend on various factors in the interview workflow.

\subsubsection{\textbf{(AI) Assistant-generated Follow-up Questions are Good: Depth-Oriented Enhancement}}
\paragraph{\textbf{Within Boundaries of Topics}}
Overall, participants generally considered (AI) assistant-generated follow-up questions to be valuable, which was reflected in two main aspects, ``supplementation'' and ``pushing the topic deeper''.

For ``supplementing'', participants emphasized that AI could bring up the finer details they had overlooked or hadn't had time to elaborate on, and continue probing them further.

\begin{quote}
    ``\textit{[AGQs] captured the core of the issue... exactly the point I had missed... it was very good, providing a supplement}.'' (P8-N)%
    
    ``\textit{When moving on to the next question, it also fully takes into account the ideas from your previous question before continuing to probe further. I feel this is a kind of complementary process.}'' (P6-N)
\end{quote}

While some mentioned how AI's follow-up questions could be ``supplementing'' during the interview process, others also talked about the aspect of ``pushing the topic deeper''. For example, P16 and P11 mentioned that the questions often pull originally broad concepts down to a more specific level to probe into the details.

\begin{quote}
    ``\textit{...for instance, I proposed a concept, and then [AI] might add it to the protocol and ask some more specific questions.}'' (P16-N) 
    
    ``\textit{I feel that the questions it asked were based on some of your previous answers, perhaps things I had overlooked, or questions framed within a certain structure that were then further expanded on. I think that's pretty good.}'' (P12-I)
\end{quote}

It is worth noting that some participants set a prerequisite for defining AGQs as useful: they are useful only when tightly aligned with the interview topic. In other words, during the interview process, the relevance of AGQs to the research theme, particularly by incorporating context, themes, and surrounding information, is one of the most fundamental preconditions; under this condition, deeper probing becomes more valuable.

\begin{quote}
    ``\textit{...the follow-up questions that were just added are very helpful for me. They built on what I originally had and went further, which allowed me to gain a more detailed understanding of the interviewees' thoughts.}'' (P6-N)
\end{quote}

The results showed that participants generally felt AGQs performed well in maintaining thematic relevance (P1-I, P2-N, P3-I, P4-N, P5-N, P6-N, P8-N, P9-I, P10-N, P11-A, P13-N, P15-A, P16-N, P17-I). For example,

\begin{quote}
    ``\textit{The good thing is that last follow-up question—it's something I didn't think about, and it really extended the issue... [at the same time] it was highly relevant to the theme of my interview study.}'' (P13-N) \\

    ``\textit{I feel it is quite relevant… [the AI-generated questions] are part of my [interview] framework.}'' (P2-N)
    
\end{quote}

\paragraph{\textbf{Out of Scope}}
While many participants emphasized the importance of close topical alignment, others adopted a more flexible perspective. For them, the defining characteristic of a useful follow-up question was not strict adherence to the interview guide, but whether it constituted a ``good question'', i.e., one that was surprising, insightful, or pushed the conversation into greater depth. 

\begin{quote}
    ``\textit{Relevance isn't really the criterion I use to judge whether a question is good or not... Asking a relevant question isn't actually that difficult... Only when it asks a question that I find particularly striking does it show that, on a deeper psychological level, its understanding of the issue is similar to mine... If it can just ask some relevant questions, I feel that's dispensable for me.}'' (P3-I)
\end{quote}

Such questions, though sometimes diverging from the interviewer's intended direction (i.e., out of scope), were experienced as useful in more exploratory or open-ended interviews, where serendipitous directions could lead to new insights.

\begin{quote}
    ``\textit{This one is more like a way to spark further ideas... It can't be said to be the most perfectly aligned with the topic, but I think it's a really good question... Overall, I feel it's quite good.}'' (P16-N)
\end{quote}

P9 vividly described this experience with the metaphor of a ``crossroad'': 
\begin{quote}
    ``\textit{When reaching a crossroad… one of its (AI's) questions could show me another path to take... and continuing along it could actually lead to some pretty good answers.}'' (P9-I)
\end{quote}

However, this precondition suggests that once the main thread is deviated from, the follow-up questions may be seen as ``disrupting the interview flow'' or ``rewriting the main purpose of the interview''. In this way, even when a follow-up was inherently a ``good question,'' it could feel misaligned if it diverted the conversation away from the direction they wished to pursue. For example, P1 mentioned feeling their questioning approach had been ``interrupted'' when AI made a suggestion:
\begin{quote}
    ``\textit{At that time, I wanted to see... where the gaps were in smart homes (the interview topic chosen by the participant), but if we shift more toward the emotional value side, then we'd probably have to talk about AI agents, and that would drift away from the original question itself...[When Wizard-AI] asked a question, my next question was already in my head... I was just interrupted, and had to make another spontaneous response and adjustment.}'' (P1-I) 
\end{quote}

Fortunately, participants quickly adapted to such dynamic processes. P1 added,

\begin{quote}
    ``\textit{It wouldn't [throw off my train of thought]. In fact, the moment [Wizard-AI] asked the question, I was already making a decision, whether to leverage it to my advantage, or to stick to my own line of thinking.}'' (P1-I)
\end{quote}

In this case, P1, with an intermediate level of expertise, can quickly adapt to the process and control the scope. 
\paragraph{\textbf{Concerns about Potentially Harmful Content}}
Though participants recognized the usefulness of AGQs, they also expressed concerns about potentially harmful content produced by AI, such as offensive remarks, aggressive responses, and generally embarrassing questions (P2-N, P15-A, P16-N, P17-I). These risks were often related to cultural background (P16-N) and to gender, age, and personality traits (P2-N, P15-A). For example, P16-N noted:
\begin{quote}
``\textit{If the AI asks something offensive, for example, criticizing someone's accent, it could really hurt the feelings of my participant. As an immigrant myself, I would not know how to handle such situations.}'' (P16-N)
\end{quote}

Nevertheless, some participants believe that the risks could be controlled in practice (P2-N, P17-I). 
\begin{quote}
``\textit{Real-life interview situations are often complex, and there will always be nuanced cases. But I think the key for me is to minimize such risks, and the IRB (Institutional Review Board) approval in advance is necessary.}'' (P17-I)
\end{quote}

According to P17, with an intermediate level of expertise, researchers as lead interviewers could to some extent control such risks if they are well-trained ethically.

Overall, participants perceived AGQs as valuable as they could supplement the overlooked parts of the conversations and facilitate the topic more deeply. While most participants regard such value as based on topic relevance, others regard thought-provoking questions as valuable, though they diverged from the original scope. However, the participants raised concerns about the potentially harmful or inappropriate AI-generated content and highlighted the ethical training and safeguards.

\subsubsection{\textbf{AGQs are Good: Making Connections, Creating Thinking Space, and Governing Pace for Human Interviewers}}

In the previous subsection, we presented the usefulness of AGQ from the content level, reliably supplementing gaps and pushing topics deeper. In what follows, we examine how lead interviewers perceive the usefulness of AGQs in relation to conversational flow. Participants described two recurrent roles during the interview workflow: (1) a connector/springboard that bridged turns and carried momentum, and (2) a buffer/holding space that created brief room to think or unblocked stalls. In addition, they highlighted pace and timing as conditions under which brief insertions were accepted and even preferred.

\paragraph{\textbf{AI-generated Follow-up could be a connector/springboard}} Participants emphasized that insertions were welcomed when they linked what had just been said to a coherent next step and felt naturally placed, often giving the human interviewer a moment to plan the subsequent probe or next move.

\begin{quote}
    

    ``\textit{On the other hand, it was continuing this conversation... I felt as if I were on a springboard... [Wizard's] question might be the second level... I wanted to step on this board and keep jumping upward...}'' (P1-I)\\
    ``\textit{It played a relatively connecting role... its question might just be a supplement to my previous question, but it really helped me think of my next question.}'' (P8-N)\\
    ``\textit{After [Wizard] finished asking such a question, what I might think is I would consider whether the question [Wizard] asked makes sense. If I feel it makes sense, then I would follow what [Wizard] said and go on to further.}'' (P10-N)
\end{quote}

In these cases, participants felt that the AI-generated follow-up can help with maintaining the momentum, connecting ideas and keeping the conversations go deeper or broader.

\paragraph{\textbf{AI-generated Follow-up could be a buffer/holding space}} At decision points (when interviewer needs to ask a follow-up question), if the interviewer felt stuck, AI proposed short, well-timed follow-ups could be like a buffer: giving the human interviewer more ``space for thinking'' (P10-N), unblocking the path, or offering an alternative direction for continuing the interview flow. For example, both P9 and P5 highlighted that the AGQs can provide directions when they get stuck and need time to think:

\begin{quote}
    ``\textit{Sometimes, its question might be a bit different from the point I was originally... thinking about, so then, I have to first stop my own thinking, and follow its pace to go in that direction. But sometimes maybe I just happened to get stuck... one of its questions could then inspire me that there is a path.}'' (P9-I)\\
    ``\textit{ It also could give me a bit of thinking time, actually … when you were answering its question, I was catching my breath… thinking what I should ask next.}'' (P5-N)

\end{quote}

\paragraph{\textbf{AGQs could timely Balance Pace}} Participants stressed that the timing of insertions was negotiable when AGQs were perceived as useful to maintain the conversation flow. Brief and high-quality insertions were acceptable, even desirable, when they kept the conversation moving forward and avoided awkward backtracking. For example, P10 specifically mentioned that the AGQs could help them dig deeper for a specific topic, and P5 highlighted that the real-time insertion of relevant follow-up questions could avoid retrospection later:

\begin{quote}
    ``\textit{It [the AGQ] might have slightly disrupted my (interview's) pace, but I felt it was okay, and I could explore this point from another perspective and dig deeper... I feel it could, to some extent, ease my concern about not digging into the topic and provide a third perspective.}'' (P10-N)
    
    ``\textit{I feel this kind of thing (jumping in for a question) is very normal... if it did not insert [questions] at that moment ... and later pulled you (interviewee) back again to talk about this... it would feel very awkward.}'' (P5-N)
\end{quote}

In summary, participants highlighted the process-oriented support of the AI-generated follow-up questions as a way of maintaining the conversational flow. Specifically, it could act as a connector/springboard that facilitates topic development and provides alternative directions. It could also be considered as a buffer/holding space, leaving brief temporary moments for reflection and planning. Additionally, participants stressed the significance of appropriate and balanced timing of insertion to help maintain the conversation flow.

\subsection{RQ2: Evaluating the Support From the (AI) Follow-up Assistant} 

In Section~\ref{sec:value}, our findings show the perceived usefulness of AGQs in conditional contexts. Beyond this, we also discovered how the lead interviewers perceive the support from the (AI)follow-up assistant. In this study, we used an AI-driven/Reverse Wizard-of-Oz method, where the researcher acted as a ``co-interviewer'' to play the role of the assistant. Because of this, participants gave feedback mainly about the assistant's behavior and timing, not about the technical capabilities of any specific model itself. Overall, participants often mentioned that the AI co-pilot could reduce the immediate cognitive load and provide emotional support, though timing is a weak spot perceived by the lead interviewers. We also discussed how the aspect of human gatekeeping could potentially influence the perceived autonomy and agency for both lead interviewers and interviewees. Lastly, we explored how trust in the (AI) follow-up assistant could be influenced by domain expertise and research question alignment.

\subsubsection{\textbf{Reducing the Cognitive Load of the lead Interviewers}} 
 In collaborative interview, most participants acknowledged the (AI) follow-up assistant's support in reducing their cognitive load, including reducing the workload and fatigue (P6-N, P12-I), improving the interview flow and time management (P7-N, P8-N), as well as capturing key information (P9-I, P14-N). Firstly, participants noted that the assistant could help ease their work when they were ``\textit{tired of asking the question}'' (P12-I) or ``\textit{the language system cannot process well immediately after getting up.}'' (P6-N)
The (AI) follow-up assistant is also perceived as an external facilitator that helps with managing the interview flow and time, particularly when the conversation risks going off topic or exceeding time. For example, P7 described how their reminders supported the interview process:
\begin{quote}
    ``\textit{The [wizard's] reminders were crucial for both me and the interviewees to return to the track, signaling when the conversation was out of the boundary or moved beyond the scope.}'' (P7-N)
\end{quote} A few participants also highlighted the (AI) follow-up assistant's support in identifying and documenting key information that they might have missed. As P9 said, 
\begin{quote}
    ``\textit{It [the wizard] can also help me capture some key information from the interviewees, including some information that I might hear but miss in that moment. I am very much looking forward to this in the future.}'' (P9-I)
\end{quote}

Additionally, participants shared their preferences for the future design of the assistant in terms of reducing the cognitive load (P1-I, P10-N, P12-I, P13-N, P14-N). Because of the difficulties in processing and selecting AGQs presented during the interviews, some participants proposed solutions such as keywords in text form (P13-N), visual cues highlighting the priority of information (P1), and involving another human co-researcher to facilitate the process (P10-N).

\subsubsection{\textbf{Providing Emotional Support for the lead interviewers}} 

In our experimental scenario, a few participants highlighted the emotional support from the (AI) follow-up assistant(P3-I, P8-N, P9-I, P12-I). They emphasized that the presence of the assistant brings about a feeling of confidence and security during the interview. As P8 noted, with the assistant's support,
\begin{quote}
    ``\textit{I would feel more assured and bold to ask my questions... a bit more reassured}'' (P8-N).
\end{quote} P10 also explained that when they could not react in time, or when they struggled to handle assertive interviewees, the assistant's follow-up brought up crucial issues spontaneously, which ``\textit{gave me a strong sense of security and reassurance}'' (P10-N).

\subsubsection{\textbf{Timing as a weak spot}}\label{sec:timing}
Most participants pointed out that the (AI) follow-up assistant was weak in handling the timing of inserting AGQs (P5-N, P13-N, P14-N, P15-A, P16-N). Some complained that the assistant tended to ask questions before the human interviewer was ready. Such early insertion disrupted the logical flow carefully designed by the interviewer. For example, P13, as a novice researcher, described how the assistant's early intervention disrupted the logic flow well prepared: 
\begin{quote}
    ``\textit{The [Wizard] actually brought up a question that I had planned for later… its pace was a bit too fast. I was intending to guide the conversation step by step, but it jumped straight to the main question. Once that was asked, there was basically nothing left to talk about}'' (P13-N). 
\end{quote}

In other situations, participants found it challenging to keep up with the AGQs proposed by the assistant and to bring the conversation back to their own track. P11, as a researcher with advanced level, also complained about the negative disruption on the conversation flow: 
\begin{quote}
    ``\textit{I remember only once when I managed to keep up and continue asking, but most of the time I couldn't. My own pace was cut and I had to skip my next question.}'' (P11-A)
\end{quote}

Notably, this resistance was more pronounced among advanced interviewers, who entered with a firmer sense of their intended trajectory; for them, an insertion that broke pacing was less easily absorbed than for novices, who more often welcomed it as scaffolding.

In addition, some participants also had trouble with losing their sense of control over timing because they did not know when the next question would come up. For example, P14 expressed confusion about the timing of the assistant's interventions during the conversation:
\begin{quote}
    ``\textit{I was unsure about the timing, and I had no idea when I should ask my next question if [the wizard] cut in. For example, after I asked something and you finished answering, I wasn't sure if the [Wizard] still had something to add, so I would usually wait a bit.}'' (P14-N)
\end{quote}

In response, some participants preferred the (AI) follow-up assistant to insert questions in the middle or later stages of the interview. As P16 put it:
\begin{quote}
    ``\textit{Just don't insert while I'm still talking. Wait until I finish, and then ask in the gap between turns. I think mid or later in the interview is better, because at the beginning we're mostly laying the ground}'' (P16-N).
\end{quote} 
P5 similarly suggested that mid-interview insertions could provide them with valuable thinking space: 
\begin{quote}
    ``\textit{I actually prefer it to insert in the middle, because when it was asking you a question and you were answering, I was already thinking about what my next question should be… so it could actually give me a bit of time to think.}'' (P5-N)
\end{quote}

In summary, we discussed the (AI) follow-up assistant's perceived role in reducing cognitive load and providing emotional support for the lead interviewers. However, the participants critically highlighted the assistant's weak point of in handling the timing of the questions and suggestions of the preferred timing. The following subsections further examines the aspects of human gatekeeping and trust in shaping the perception on the (AI) follow-up assistant.

\subsubsection{\textbf{Human Gatekeeping and Control}}
\paragraph{\textbf{Gatekeeping as Autonomy and Responsibility}}
In collaborative interview, most of the participants perceived human researchers as the ultimate gatekeepers over the questions (P2-N, P6-N, P7-N, P13-N, P14-N, P17-I). Researchers keep the ultimate authority to decide whether to adopt AGQs and to adjust them for appropriateness and relevance to the research focus. They emphasized that keeping such authority was a fundamental right of the interviewer. 

\begin{quote}
    ``\textit{I think I should have the basic right to decide on the spot, like, I can choose not to ask a question, or choose when to ask it.}'' (P7-N)
    
    ``\textit{I'd probably look over what it was going to say first. I wouldn't let it just speak out directly}'' (P2-N).
\end{quote} In the future design, a few participants suggested that the assistant could, for example, ``\textit{turn on a light}'' (P1-I) (also be mentioned by P12-I) or ``\textit{raise its hand}'' (P6-N) before authorizing them the opportunity to select from the proposed follow-up questions.

Some participants highlight this type of autonomy by emphasizing the role of the (AI) follow-up assistant as a research tool rather than a human requiring high effort to communicate with (P10-N, P13-N, P17-I). As P13 participant noted, 
\begin{quote}
    ``\textit{If [the(AI) follow-up assistant] proposes a question that you think is inappropriate, you can directly tell it to change. But with humans, they might insist on their point of view, which I guess is more time-consuming.}'' (P13-N)
\end{quote}

In these cases, we found that the participants value their lead position as the gatekeeper and their authority over the (AI) follow-up assistant.

\paragraph{\textbf{Gatekeeping as Respect for Human Agency}}
Additionally, human gatekeeping can be framed as a way to respect interviewees, safeguarding their agency, boundary and comfort while ensuring alignment with ethical guidelines (P1-I, P7-N, P10-N, P12-I, P14-N). Potential disrespect and discomfort for the interviewees with the presence of the (AI) follow-up assistant were highlighted. For instance, P7 reflected from the perspective of an interviewee interacting with the assistant and commented:
\begin{quote}
    ``\textit{I just felt that mode of AI-to-Human is very disrespectful to me. It is really a matter of respect...it should be either human-to-human or AI-to-AI. It's just the basic respect.}'' (P7-N)
\end{quote} Another participant expressed concerns about losing control over configuring the personality of the (AI) follow-up assistant, which could potentially cause discomfort for interviewees. 
\begin{quote}
    ``\textit{If its personality cannot be adjusted and it turns out to be aggressive and cause conflict, I am very concerned that it might violate the boundaries of the interviewee.}'' (P1-I)
\end{quote}

The interviewee's right to be informed is particularly emphasized when working with vulnerable populations (P7-N, P10-N, P12-I). For instance, one participant (P10-N) who worked with children experiencing depression highlighted concerns about the feasibility of using the AI follow-up assistants in such contexts. 
\begin{quote}
    ``\textit{Normally, it is difficult for me to gain the trust of interviewees and have them participate in the interview. Things would be more complicated if they were asked to engage with a third party[(AI) follow-up assistant]}'' (P10-N). 
\end{quote}

In summary, the findings suggest that human gatekeeping is viewed as a critical mechanism. It served not only to preserve researchers' autonomy and decision-making authority, but also to protect and respect interviewees' agency, comfort, and personal boundaries during the interview process.

\subsubsection{\textbf{Domain Expertise and RQ Alignment Build Trust}}
\paragraph{\textbf{Domain Expertise as a Basis for Trust}}
Most participants believed that domain expertise was a fundamental basis for establishing trust in co-interviewers. In our experiment, some participants (mostly novice researchers) acknowledged the co-interviewer's domain expertise before we disclosed the role of the AI-driven wizard. For instance, P16 noted that the co-interviewer's questions were of such quality that [the wizard] could ``reach the doctoral level in my field'' (P16-N). To some extent, it demonstrated how the (AI) follow-up assistant could gain the interviewers' trust with the domain expertise.

On the other hand, co-interviewers without sufficient background knowledge were not regarded as qualified collaborators, particularly by participants as intermediate researchers. As P15 explained, 
\begin{quote}
    ``\textit{When I was trying to dig deeper into the question, [the wizard] inserted a question that redirected the discussion to a more general level. That was the time I realized [the wizard] might not be familiar with our field}'' (P15-A).
\end{quote} For a stronger academic foundation of the (AI) follow-up assistant, participants further suggested that equipping the assistant with domain knowledge and fine-tuning it would improve its reliability (P10-N, P11-A, P17-I).

\paragraph{\textbf{Research Question Alignment in Building Trust}}
In addition to the domain expertise, participants also highlight the significance of Research Question (RQ) alignment in building trust. They expressed a preference for follow-up questions that were closely aligned with their RQ to build trust in the (AI) follow-up assistant (P1-I, P11-A, P12-I, P14-N, P16-N, P17-I). For example, P16 highlighted that the trust in the wizard remained unchanged after learning that it is AI-driven, as the follow-up questions were well aligned with their research topic: 
\begin{quote}
    ``\textit{I don't really mind as long as [the assistant] can ask follow-up questions closely related to my research topic and within the scope.}'' (P16-N).
\end{quote}

Conversely, insufficient alignment with the RQ can undermine the perceived credibility. As P11 noted, \begin{quote}
    ``\textit{Maybe because [the wizard] didn't know about my research questions, some of the follow-up questions [the wizard raised] were off topic, though they were still good...But I did not want to adopt them.}'' (P11-A)
\end{quote} In response, for future design of the (AI) follow-up assistant, a few participants suggest providing their interview guide or research background in advance to ensure better alignment with their research context (P11-A, P17-I).

Across these accounts, participants regard the domain expertise and RQ alignment as two key factors in building the trust on the (AI) follow-up assistant. Trust was built and strengthened when the assistant could demonstrate domain expertise, sufficient background knowledge and a strong alignment with the research topic and context.

\subsection{Findings Summary}
Across both research questions, three findings stand out. First, AGQs were perceived as useful conditionally: they reliably supplemented overlooked points and pushed topics deeper (RQ1), but their value depended on topical relevance, well-judged timing, and human ethical oversight. Second, usefulness was as much a matter of conversational mechanics as of content, AGQs were valued as connectors/springboards and as buffers for thinking, and were rejected when their timing disrupted the interviewer's planned flow. Third, participants consistently positioned the human as the ultimate gatekeeper, both to safeguard research focus and to protect interviewees' agency, and grounded their trust in the assistant's domain expertise and alignment with their research questions. Finally, expertise shaped these perceptions: novices tended to treat AGQs as guidance and reassurance, whereas advanced interviewers applied a higher bar, valuing suggestions only when they were both surprising and clearly advanced their own line of inquiry. These patterns inform the design implications discussed next.

\section{Discussion}

In this study, participants' acceptance of AI suggestions depended not only on content quality but also on the perceived alignment with human values~\cite{10.1145/3442188.3445922} and interview goals. This echoes broader discussions in HCI about how human trust in AI is mediated by perceptions of fairness, accountability, and control~\cite{shneiderman2020human}. Interviewers emphasized that they must retain ultimate authority over the interaction, both to preserve research integrity and to maintain rapport with interviewees.

We clarify that our focus on AGQs is not intended to reduce qualitative interview to question generation, nor to suggest that interview can be fully automated or replaced by question generation alone. Instead, we treat follow-up questions as an important and observable entry point into the interview process because they partially materialize how interviewers interpret participants' responses, maintain the interviewer–interviewee relationship, manage the unfolding trajectory of the interview, and make situated ethical decisions about what can be asked, when, and how. Meanwhile, while novice interviewers tend to view AGQs as a source of guidance and confidence, more advanced interviewers were more likely to emphasize the importance of developing their own interpretive judgement and questioning strategies. This highlights a potential tension between providing support and fostering expertise, which could be further discussed in future research.

\subsection{Tensions and Concerns of Introducing AI into Qualitative Interviews}

While our findings point to concrete benefits, we take seriously the position, held by many qualitative researchers, that introducing AI into data collection carries real risks along interactional, methodological, and societal dimensions. Interactionally, even well-aligned AGQs can fracture the rapport and turn-taking on which disclosure depends, as several participants experienced when timing was off. Methodologically, an over-reliance on machine-generated probes risks narrowing inquiry toward what the model finds salient and may, for novices, inhibit the slow development of interpretive judgement that interviewing expertise requires, a tension our advanced participants implicitly voiced. Societally, normalizing AI in interviews raises questions of consent, accountability, and the gradual de-skilling of researchers. We do not regard these concerns as resolved by our design; rather, we treat them as boundary conditions, and argue that AI should be confined to a bounded, language-based sub-task under continuous human authority, not extended to the relational and ethical core of interviewing.

As described in Section~\ref{subsec:study-procedure}, our design involved temporary non-disclosure of the AI's involvement, followed by full disclosure and a reflective debrief within the same session. This arrangement warrants explicit ethical reflection. We adopted this minimal, time-limited deception because foreknowledge of AI authorship would have introduced reactivity (Hawthorne~\cite{adair1984hawthorne}/Halo effects~\cite{pohl2016cognitive,nisbett1977halo}) that directly confounds our research questions about perceived quality and trust. We judge the ethical cost to be low: no sensitive personal data were elicited, and participants were debriefed within the same session. We nonetheless distinguish this experimental arrangement from real-world deployment. In practice, AGQs delivered to actual interviewees raise unresolved questions of who bears accountability for an AI-suggested question, how participants should be informed of AI involvement, and how trust is renegotiated once disclosed. We therefore recommend that real-world use foreground transparency and locate ultimate responsibility with the human interviewer, rather than replicating the non-disclosure used here for experimental control.

\subsection{Design Implications for Future Human-AI Collaborative Tools and Practical Suggestions}

\subsubsection{\textbf{Question Quality vs. Communication Skills in Semi-Structured Interviews}}\label{discussion:skills}

Our study suggests that the value of AGQs cannot be adjudicated as a simple either-or choice between question quality and communication skills. Rather, they operate as coupled conditions. On the content dimension, AGQs consistently supplemented missed sub-dimensions, pushed broad notions to analyzable specifics, and acted as connectors or springboards across turns. This was especially critical for novices who struggle with split-second probing and with perceiving subtle cross-references in talk~\cite{pope1974experienced, MYERS20072}. In this sense, a good question is a necessary foundation for depth. Yet interview is first and foremost a social practice~\cite{roulston2010,Magaldi2020,adams2015}: pace, tone, and turn-taking stability decide whether a question can land without threatening rapport or shutting down disclosure~\cite{whiting2008semi,marshall2014}. When the foundation is unstable, even high-quality probes are experienced as interruptions; when the foundation is stable, interviewers appropriate even modest suggestions into deeper inquiry.

This interactional mechanism aligns with the structure-with-flexibility ideal of semi-structured interview and with grounded, improvisational probing \cite{charmaz2006,marshall2014}. It also explains why participants accepted AGQs when they preserved immediate conversational flow (brief, well-timed insertions) and framed neutrally, and why they rejected them when they shifted the topic or collided with the interviewer's planned trajectory. For novices, AGQs created a momentary buffer space to think, thereby reducing the cognitive load of composing the next probe while maintaining the momentum of the conversation. For experienced interviewers, who typically enter with a clearer sense of thematic scope and a practiced interview flow, the bar for acceptance was higher: they resisted scope drift and valued AGQs chiefly when the suggestion was both surprising and clearly advancing their current line of questioning.

These observations motivate a pragmatic division of labor in human-AI collaboration. In the short term, the quality of AGQs is already serviceable for scaffolding depth, while communication skills (pace, tone, timing) remain the principal bottleneck for many novices. Our AI-driven/Reverse Wizard-of-Oz co-interview design operationalizes this: by routing AGQs backstage to a human co-interviewer, the system mitigates current limits in machine prosody and turn-taking, lets novices practice foundation management and insertion timing with reduced pressure, and preserves the ethical tenor of the exchange \cite{Kincaid1957a,velardo2021,jiang2021supporting}. Over time, as novices master interview skills, they can graduate to the lead interviewer role and use AGQs to broaden perspective and test boundaries without sacrificing flow. Conversely, for experienced interviewers, AGQs should default to backstage, with steerable scope (ranging from on-topic to exploratory), typed probe options (clarify, example, contrast, causal, boundary), and lightweight acceptance controls; only exceptional, high-gain suggestions warrant frontstage micro-insertions.

In summary, question quality and communication skills are co-requisites rather than substitutes. Designing AGQ tooling as ``from stabilize the foundation to deploy precise probes'' integrates long-standing guidance-improvisation tensions in semi-structured interviews \cite{roulston2010,Magaldi2020} into a concrete interaction contract. This reframing helps explain the mixed reactions to AGQs observed in practice (at times accepted, at times rejected) and, more importantly, highlights \textbf{a training pathway for novice interviewers}: initially offloading question generation to reduce cognitive burden and allow practice in pace and tone, then, once interactional stability is achieved, strategically using AGQs to expand both depth and scope of inquiry.

\subsubsection{\textbf{Modes of Interaction: Timing and Modality of Suggestions}}
Our findings demonstrate that the perceived usefulness of AGQs was shaped not only by the quality of the questions themselves, but also by the mode of delivery and the timing of their insertion. Participants repeatedly noted that small, well-timed interjections could preserve or even enhance conversational flow, whereas poorly timed interruptions risked breaking flow and drawing attention away from the participant's narrative. These observations resonate with prior work on semi-structured interviews that highlights silence management and pacing as critical interviewer skills~\cite{price2002,pope1974experienced}. 

\begin{figure*}[htbp]
    \centering
    \includegraphics[width=0.9\linewidth]{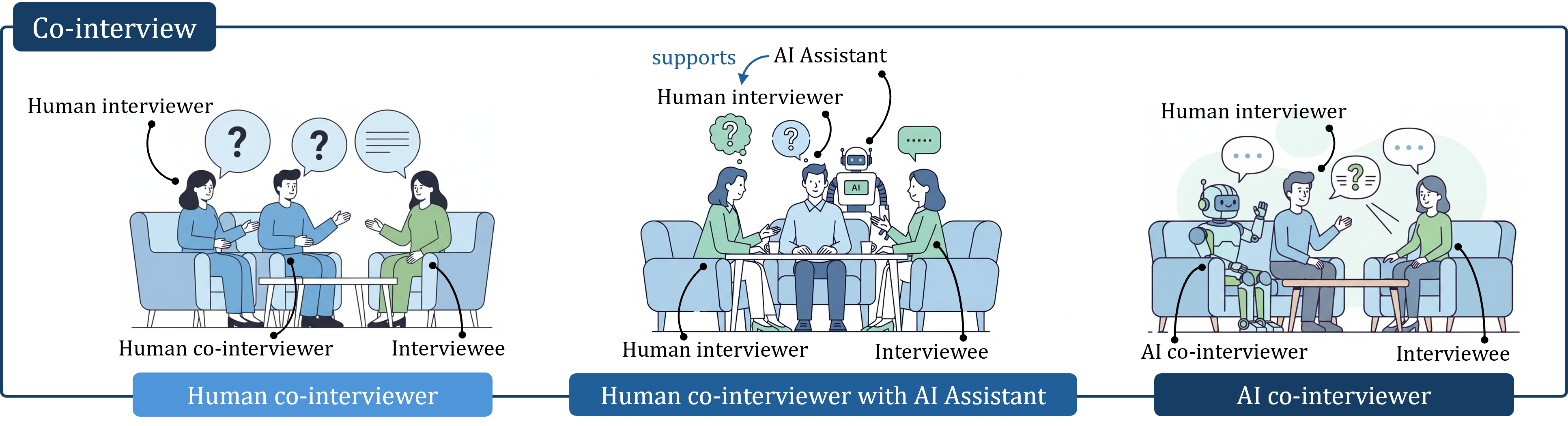}
    \caption{Co-interview configurations. The figure illustrates three configurations of co-interview settings. (1) Human co-interviewer: two human interviewers collaborate in questioning the interviewee. (2) Human co-interviewer with AI co-pilot: a human interviewer is supported by an AI system that generates suggested follow-up questions. (3) AI co-interviewer: the AI directly acts as a co-interviewer alongside the human interviewer when interacting with the interviewee.}
    \label{fig:co-interview}
\end{figure*}

\begin{figure*}[htbp]
    \centering
    \includegraphics[width=0.9\linewidth]{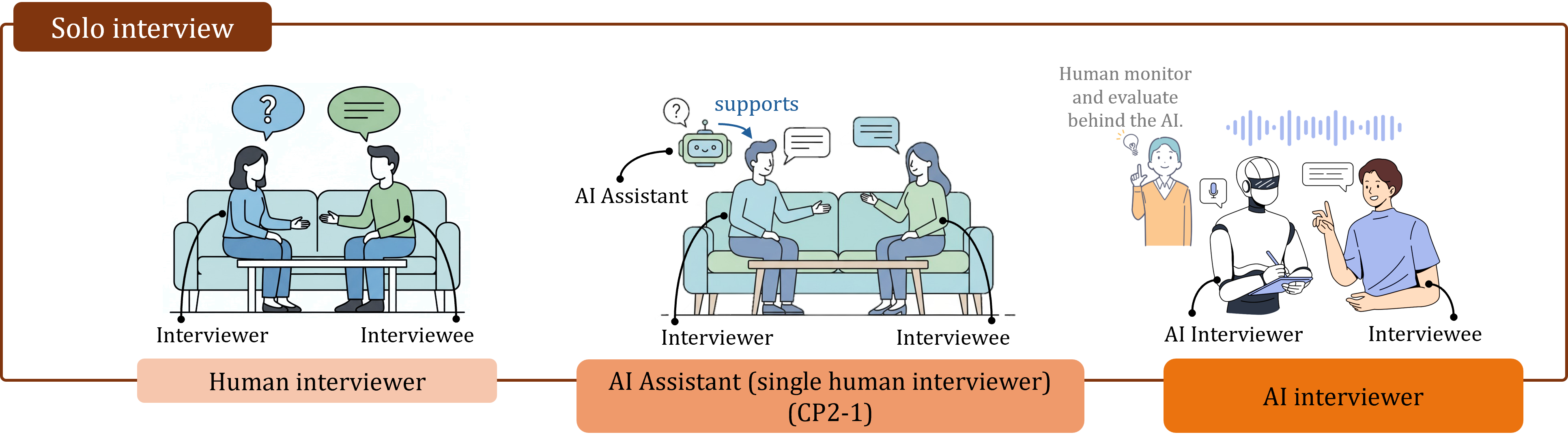}
    \caption{Solo interview configurations. The figure illustrates three configurations of solo interview settings. (1) Human interviewer: a single human interviewer interacts directly with the interviewee. (2) AI co-pilot (single human interviewer): a human interviewer is supported by an AI system that provides suggested follow-up questions or guidance. (3) AI interviewer: the AI takes over the interviewer role and engages directly with the interviewee while the human interviewer worked in the backstage to monitor the interview}.
    \label{fig:solo-interview}
\end{figure*}

To clarify these design tensions, we differentiate two overarching collaboration paradigms, \textbf{(CP1) AI-supported co-interview} (see Figure.~\ref{fig:co-interview}), and \textbf{(CP2) AI-supported solo-interview} (see Figure.~\ref{fig:solo-interview}).


\begin{figure*}[htbp]
    \centering
    \includegraphics[width=0.7\linewidth]{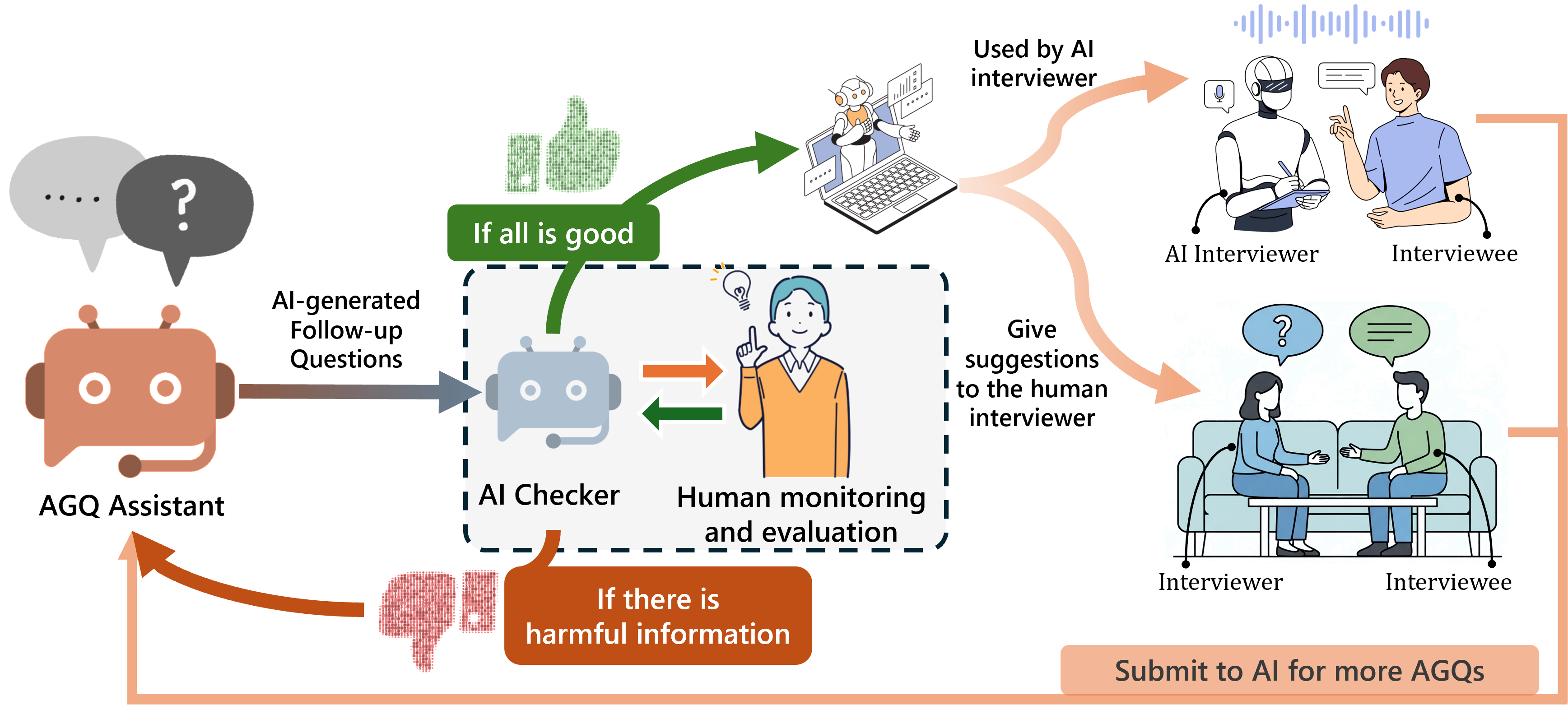}
    \vspace{-0.2cm}
    \caption{AI Checker mechanism. The figure illustrates the proposed AI Checker, designed to enhance the safety and reliability of AGQs. The AGQ Assistant first produces candidate questions, which are then evaluated by the AI Checker. If the content is safe, the questions are passed on for use by either a human interviewer or an AI interviewer. If harmful or inappropriate information is detected, the questions are filtered out and redirected for revision before being reused.}
    \label{fig:aiChecker}
\end{figure*}

Building on these paradigms, we further propose an automation strategy (see Figure.~\ref{fig:aiChecker}) that introduces AI into qualitative interview while explicitly safeguarding ethical standards such as human authority and participant consent.

In the paradigm of AI-supported co-interviewer, our experimental results and participants' proposed design suggestions brought out two concrete sub-interaction modes, namely \textbf{(CP1-1)} setting AI as backstage support for the secondary human interviewer, and \textbf{(CP1-2)} frontstage direct interaction. In the paradigm of AI-supported solo-interview, based on the traditional one-to-one interview mode, we distinguish between two analytically useful configurations: \textbf{(CP2-1)} AI as an assistant supporting human interviewer, and \textbf{(CP2-2)} AI as an independent solo-interviewer.

\begin{figure}[h]
    \centering
    \includegraphics[width=0.96\linewidth]{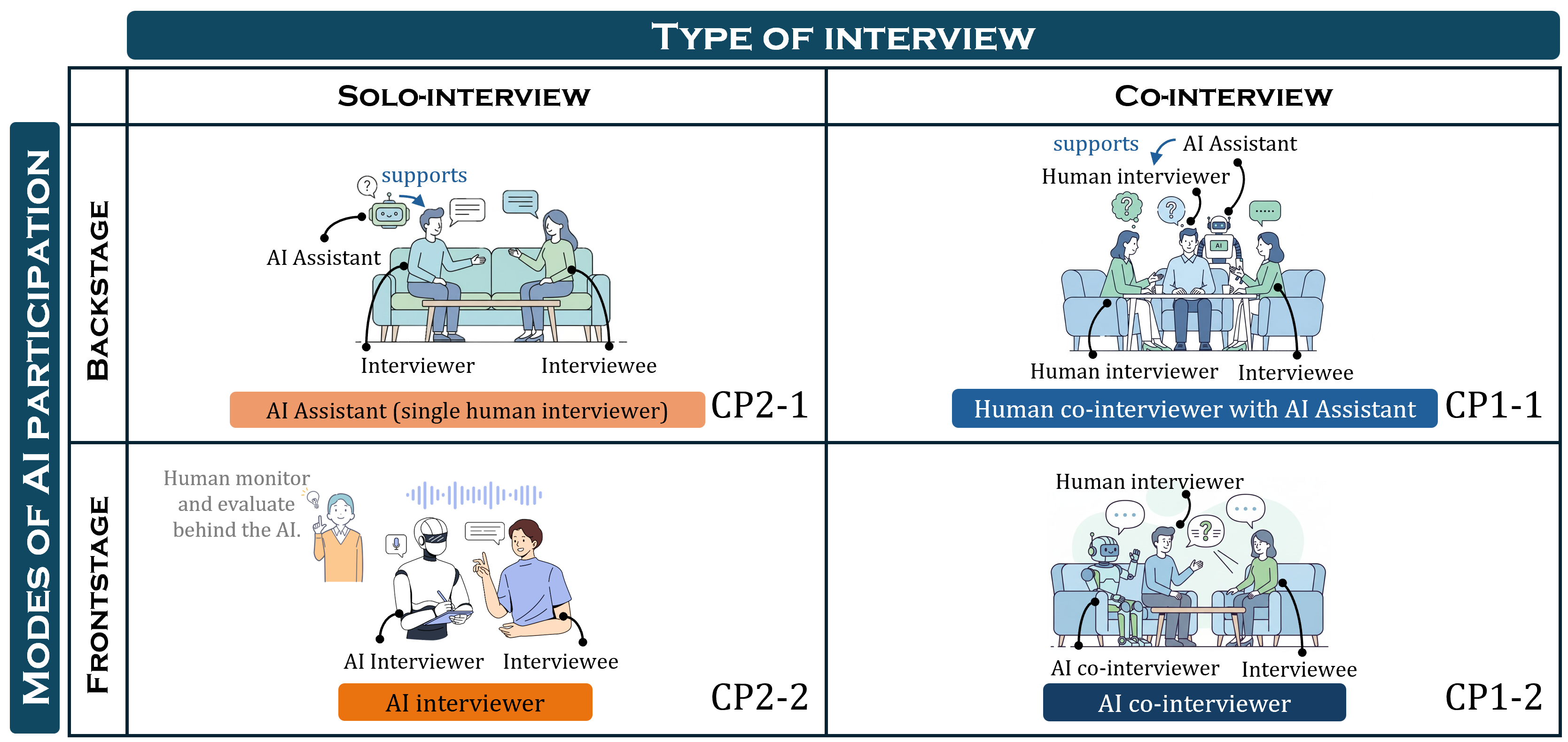}
    \vspace{-0.2cm}
    
    \caption{A Quadrant Chart for 4 Different AI-Supported Interview Modes. The figure categorizes interview configurations along two dimensions: type of interview (solo-interview vs. co-interview) and modes of AI participation (backstage vs. frontstage). The four resulting quadrants illustrate distinct AI-supported modes: (CP2-1) solo interview with an AI co-pilot operating backstage to support a single human interviewer; (CP2-2) solo interview with the AI acting as the interviewer in the frontstage while the human interviewer worked in the backstage to monitor the interview}; (CP1-1) co-interview with a human co-interviewer supported by an AI co-pilot backstage; and (CP1-2) co-interview where the AI serves directly as the co-interviewer in the frontstage.
    \label{fig:4modes}
\end{figure}
\vspace{-0.2cm}

\paragraph{\textbf{CP1-1: Backstage AI Assists Two Human Interviewers, Supporting the Secondary Human Co-Interviewer}}
First, the mode of backstage supporting the secondary human co-interviewer, who would usually support the lead interviewers to take notes and prepare follow-up questions and probes, is very similar to the research method used in this study, that is, the AI co-pilot provides AGQs for the secondary human co-interviewer ``in places invisible to participants,'' and then the secondary human co-interviewer rephrases and expresses them. In this mode, the interview itself contains the benefits of multi-to-one human interviews~\cite{monforte2021tinkering}, and novice interviewers can improve interview skills from the learning roles and interactional relationships discussed in Section~\ref{discussion:skills}. In addition, the secondary human co-interviewer still holds the final decision-making authority over AGQs during the interview process, which is of great help in reducing the spread of potentially harmful information generated by AI~\cite{10.1145/3491102.3501965,mechergui2024goal}. In other words, when the secondary human co-interviewer uses AGQs to ask questions to the interviewee, there will first be a round of subjective screening relying on human moral values and perspectives, and then the question will be expressed in a more natural way after being rephrased by humans (for example, through signals such as ``raising a hand,'' or more colloquial expressions), collaborating with the lead human interviewer to ask the interviewee. This ensures that possible ill-timed questions generated by AI will not appear in the interview process. In addition, AI should also provide further explanations of the questions and suggestions on various ways of phrasing them, in order to more comprehensively assist the secondary human co-interviewer. 

\begin{figure*}[htbp]
    \centering
    \includegraphics[width=0.8\linewidth]{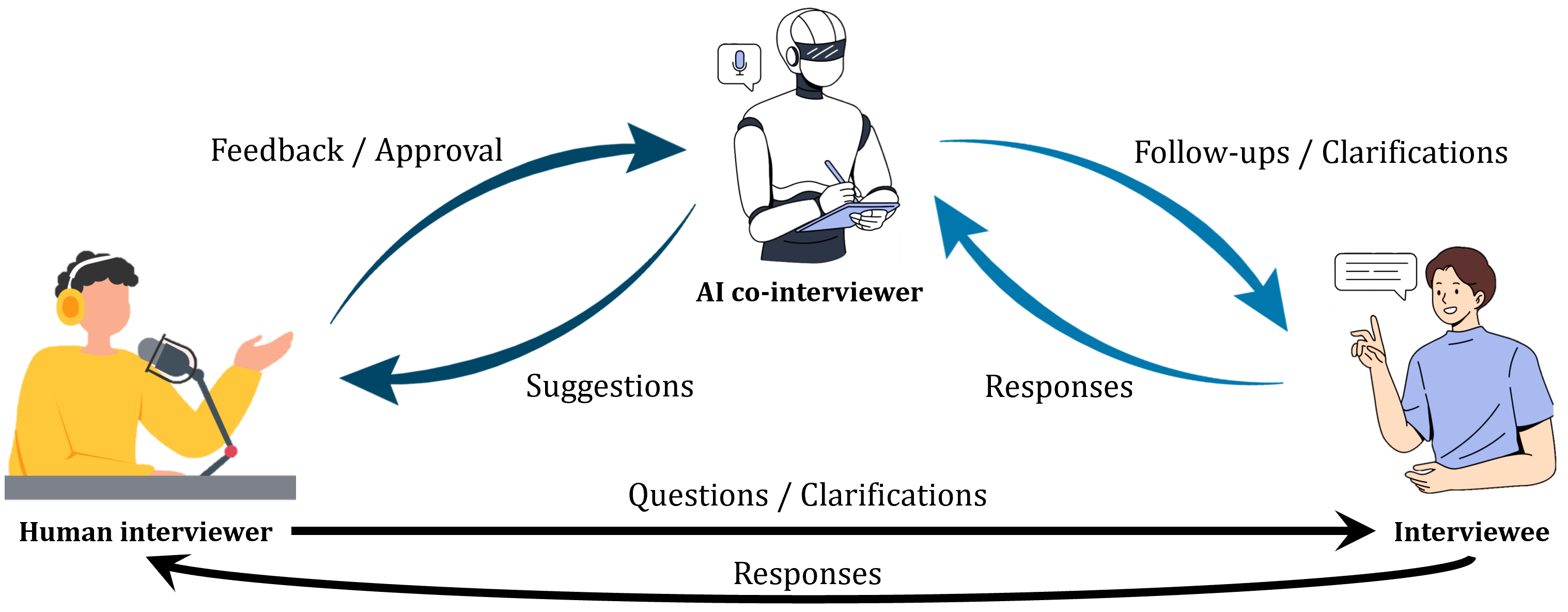}
    \vspace{-0.2cm}
    \caption{Bidirectional Questioning: Mutual inquiry among AI, human researcher, and interviewee. The AI co-interviewer provides suggestions to the human interviewer and receives feedback or approval in return, while also engaging with the interviewee through follow-ups and clarifications. The human interviewer and interviewee maintain the primary question–response exchange, augmented by AI interventions.}
    \label{fig:Bidirectional-Questioning}
\end{figure*}

Building on this backstage support paradigm, our findings also suggest that interviews may benefit from a more interactive extension, namely \textit{bidirectional questioning} (see Figure.~\ref{fig:Bidirectional-Questioning}). Instead of keeping AI solely as  passive backstage role, this approach treats AI as a reflective aid that helps human interviewers notice possible moments for follow-up or clarification. In practice, this means that AI could not only provide follow-ups and clarifications but also request additional details or verify its own understanding of participants' statements under the guidance of the human interviewer. Such active participation creates opportunities for human-led clarification and reflection around the interviewee's responses. For example, the AI may probe unclear aspects of the interviewee's response, while the human researcher can clarify or expand on AI-generated prompts. However, balancing initiative in this relationship is critical, and AI's contributions should remain complementary to the human researcher rather than competitive~\cite{krakowski2023artificial}. 

A further implication of this design is that questioning can become a site of learning and alignment on both sides. For AI, feedback from the interviewer and responses from the interviewee can serve as alignment signals, gradually improving its sensitivity to tone, pacing, and emotional resonance over time. This resonates with ongoing work on human-centered alignment, which emphasizes transparency, responsiveness, and socio-emotional attunement as prerequisites for building trust~\cite{rinne2024interpersonal}. For secondary human interviewers, bidirectional questioning also opens opportunities for professional growth. Novices, in particular, can internalize interview skills by observing and building on AI's suggestions, shifting their focus from the mechanics of generating questions to higher-order abilities such as pacing, tone, and rapport-building. Here, AGQs are not treated as substitutes for these higher-order skills, but as concrete moments through which such skills can be noticed, practiced, and reflected upon. More experienced interviewers, meanwhile, may expand their repertoire by encountering unconventional yet effective questions posed by AI. In this sense, AI is not only a backstage assistant but also a reflective partner that contributes to both immediate interview performance and longer-term skill development.
\vspace{-0.1cm}

\paragraph{\textbf{CP1-2: Frontstage AI co-pilot as a Co-Interviewer and Work Together with Another Human Interviewer}}
For the mode of frontstage direct interaction, the AI will participate in the interview as a co-interviewer; that is, it will use AGQs to ask questions to the interviewee. This way not only allows a single human interviewer to carry out interview work without needing two human interviewers, but also effectively helps the human interviewer create a buffer (thinking space) or serve as a connector and springboard. In addition to the advantage of reducing the burden on a single interviewer, as in other co-interviewer tasks, our results also show that this practice may require additional design considerations. First, expression ability and communication skills at the level of human experts are necessary (including the language used, speech, and tone). Our participants' feedback and prior studies both show that unnatural expressions (e.g., unnatural speaking flow, machine-like voice) greatly reduce participants' sense of presence and trust~\cite{wang2017tacotronendtoendspeechsynthesis, 10.1145/3382507.3418839, niculescu2013making}, and further degrade interview quality. Fortunately, at present, high-quality speech models can already achieve human-like performance~\cite {zeng2024glm4voiceintelligenthumanlikeendtoend}. Although it is very important to acknowledge and address the potential risks of such highly human-like speech (for example, fraud and hallucination)~\cite{roberts2024artificial,sun2024ai}, this does not diminish its great potential for interview tasks. 

\begin{figure*}[htbp]
    \centering
    \includegraphics[width=0.9\linewidth]{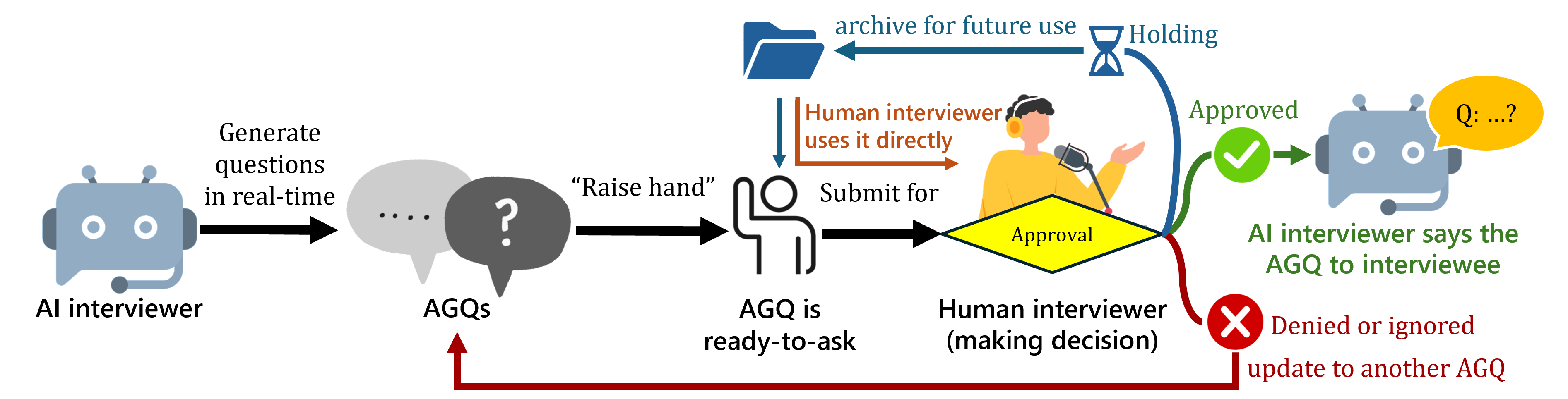}
    \vspace{-0.2cm}
    \caption{Illustration of the frontstage AI co-interviewer mode with human approval.  After generating a candidate AGQ in real time, the AI co-interviewer signals its intent to ask (e.g., through a ``raise hand'' signal).  The human interviewer then decides whether to approve, reject, or delay the question.}
    \label{fig:frontstage}
\end{figure*}

In addition to needing to improve the AI co-interviewer's expression ability to the expert level, our results also show that participants consider authority during the interview process. On the one hand, researchers (human interviewers) need to guide the research direction and pace of interview tasks, which are determined by the research questions and the researchers' concerns. On the other hand, again, some researchers, especially novice researchers, may also benefit from AI-led interview (through connector, springboard, and buffer effects). This leads to seemingly contradictory attitudes toward authority or control of the interview. However, transferring control of the AI co-interviewer's speaking turns to the human interviewer may help balance the two situations. Specifically, after the AI co-interviewer generates AGQs in real time, before asking questions, it can first signal to the human interviewer with a ``raise hand'' gesture, and then ask only with the human interviewer's permission (as shown in Figure.~\ref{fig:frontstage}). In this way, the human interviewer can minimize the possibility of their line of thought or interview flow being interrupted by inserted questions, while through the active behavior of calling the AI co-interviewer to ask, the beneficial functions of the AI co-interviewer (such as creating buffers, connectors, and springboards) are transformed into a more controllable form of dynamic support. It is also worth noting that we further suggest introducing a ``holding'' mechanism at the point of human interviewer decision-making, which can preserve those AGQs that are valuable but not yet suitable to be asked, given the progression or timing of the interview, archiving them for later use (e.g., subsequent turns or future interview rounds).

\vspace{-0.2cm}

\paragraph{\textbf{CP2-1: AI Assists a Solo Human Interviewer}}
One-to-one interviews are the main form of semi-structured interview, partly because they are easier than co-interviews in terms of staffing and organizational management (e.g., flexibility, division of labor, transcription), and thus have been widely adopted in academia and industry~\cite{adams2008questionnaires1}. However, unlike co-interviews, a solo interviewer can hardly obtain extra help or collaborative gains during the session. They must perform multiple tasks at once—tracking the interviewee's answers and reactions, taking notes, asking follow-ups, following the interview guide, and adjusting pacing, which imposes greater cognitive and attentional demands~\cite{anderson2015learning,sharma2018attentional}. This makes a solo interviewer more dependent on their own experience, communication skills, knowledge base, and on-the-spot judgment.

 Based on our findings, the AGQ Assistant can support solo interviewers by understanding the ongoing conversation, providing real-time reflective prompts, and reducing some of the cognitive demands of managing multiple interview tasks. In this setting, the AI can act as a backstage assistant that continuously provides support (similar to CP1-1). However, the cognitive and attention demand of solo interview makes the previously proposed ``raise hand'' signaling and fully displayed AGQs difficult to apply effectively in a one-to-one session. Thus, in such highly dynamic conditions, consistent with participants' feedback in our study, we recommend using \textit{AI-generated topical keywords} as cues for solo interviewer rather than full AGQs. In this mode, control remains entirely with the human interviewer: the AI, as an assistant, provides only keywords for potential follow-ups, supporting the interviewer's line of thinking rather than supplying exact wording. This design choice is motivated not only by task demands and cognitive load, but also by minimizing extraneous processing and split attention, consistent with Cognitive Load Theory~\cite{sweller1988cognitive,10.1145/1180639.1180831}.

\paragraph{\textbf{CP2-2: AI as Independent Interviewer Without Human Interviewers}} 
A more radical sub-mode within the solo-interview paradigm is to delegate the interview task entirely to the AI, allowing it to function as an independent interviewer, meanwhile the human interviewer will work backstage to mentor the process with limited cognitive load. The idea of allowing machines to conduct interviews is not new: it has been discussed in the context of early chatbots and automated survey systems~\cite{9888003}. However, it has never been as close to practical realization as it is today with the advent of LLMs. In this configuration, the AI directly poses questions, manages pacing, and follows the interview guide without continuous human mediation. This approach offers obvious advantages in scalability and efficiency, as it reduces human labor costs and enables parallel data collection across large participant pools \cite{10.1145/3722212.3725134}. It may also provide consistency in question delivery, which is difficult to achieve when multiple human interviewers are involved. However, our results and prior literature highlight significant methodological and ethical concerns. From a methodological perspective, interview quality not only depends on generating relevant questions but also on building rapport, reading subtle cues, and flexibly adapting to unexpected turns \cite{roulston2010,whiting2008semi}. Current AI systems still lack the socio-emotional sensitivity and contextual judgment necessary for such work. From an ethical perspective, participants may feel deceived or uncomfortable if they realize they were interacting with a machine rather than a human interviewer \cite{10.1145/3442188.3445922,nunamaker2011embodied,wangmo2019ethical}. The human interviewer backstage as a gatekeeper can mitigate such concerns to some extent. In addition, current AI systems still struggle with hallucinations and the generation of potentially harmful content—not only violent, sexual, or fraudulent material, but also inappropriate linguistic habits that may be unsuitable in interview settings. 

To address these concerns, we propose a multi-agent system with human in the loop, as illustrated in Figure.~\ref{fig:aiChecker}, in which an additional AI \textit{checker} is introduced between the AGQ assistant and human interviewer. Acting as a combined judge~\cite{NEURIPS2023_91f18a12}, the AI checker reviews and evaluates each AGQ to determine whether it is appropriate. If deemed suitable, the question is passed to the AI interviewer to deliver; if not, the AGQ is flagged for human review and will be regenerated. This design not only addresses methodological and ethical concerns, but also resonates with principles of value-sensitive design \cite{doorn_value_2013} and prior safety research on multi-agent language model architectures. Recent work on toxicity filtering \cite{gehman2020realtoxicitypromptsevaluatingneuraltoxic,NEURIPS2021_2e855f94} and self-critiquing or reviewer models \cite{saunders2022selfcritiquingmodelsassistinghuman} demonstrates that layered AI mechanisms can effectively reduce hallucinations and inappropriate content. By combining the human and AI checker, the system operationalizes these insights in the interview context, reducing risks while maintaining scalability. Although the AI checker has particular value in scenarios where AI conducts interviews independently, this approach can also be applied across the other modes introduced earlier, helping mitigate the risks of AGQs and reducing the burden on human interviewers. We stress, however, that the AI checker addresses only formalizable content risks, toxicity, evident inappropriateness, or topical drift, and does not, and cannot, monitor the relational, interpretive, and ethical dimensions of interviewing, which remain the irreducible responsibility of the human interviewer. We thus frame the checker as augmenting human oversight, not as substituting for it or as independently ``reducing'' relational-ethical risk.

To summarize, our exploration of co-interview (CP1) and solo-interview (CP2) paradigms illustrates how the role of AI in semi-structured interviews cannot be considered in isolation from interviewer configuration. In multi-user scenarios, AI can be embedded as a visible or invisible co-interviewer, shifting the dynamics of coordination, role allocation, and authority between collaborators. In single-user scenarios, by contrast, the interviewer shoulders all interactional and logistical demands, making cognitive load and attention management the central challenges. 

In these scenarios, AGQs show consistent value in supplementing and deepening conversation, but their mode of delivery, whether backstage or frontstage, as keywords or full utterances, determines whether they are experienced as supportive or disruptive. This comparison underscores that effective AI integration is not a ``one-size-fits-all'' solution, but must adapt to the interviewer context. For solo-interviewers, lightweight and non-intrusive scaffolding helps preserve focus, while for co-interviewers, explicit coordination protocols and approval workflows mitigate conflicts. Ultimately, the contrast between single-user and multi-user scenarios highlights a broader design principle: AI in qualitative interview should preserve human agency while strategically redistributing effort, thereby enhancing both the efficiency and the ethical integrity of human-AI collaboration.

\section{Conclusion}
In conclusion, this study employed an AI-driven/Reverse Wizard-of-Oz approach to examine the usability of AGQs and the collaborative potential of the AGQ Assistant in semi-structured interviews. Our findings suggest that interviewers generally perceived AGQs as able to improve the quality and depth of follow-up questions, and thereby to support, though not guarantee, richer inquiry, while also revealing several challenges that remain to be addressed. Building on these insights, we identified four modes of AI collaboration and proposed a bidirectional questioning framework to guide future practices in interview settings. Finally, to further enhance usability and mitigate potential risks, we introduce the concept of an AI Checker as a design direction for the next generation of AGQ Assistants aiming to align AGQ generation with value-sensitive and multi-agent safety principles.

These contributions matter because they move the conversation beyond a binary of whether AI belongs in qualitative research toward how it can be bounded responsibly. Our empirical account delineates where AGQs add value and where they fall short; our role-allocation framework gives researchers a concrete vocabulary (CP1-1 to CP2-2) for deciding when AI should sit backstage versus frontstage, and for whom; and our design suggestions, topical keyword cues for solo interviewers, raise-hand approval for frontstage co-interviewers, and a human-plus-AI-checker safeguard—translate these distinctions into actionable interaction contracts that preserve human agency while redistributing effort. Throughout, we position LLMs as complements to, rather than substitutes for, human interpretive and ethical judgement, which remains the irreducible core of qualitative interviewing.

Our study has two limitations that suggest directions for future research. First, we did not systematically capture or analyze multimodal interactional cues related to relational and emotional dynamics. Thus, our findings mainly reflect participants' retrospective accounts and our qualitative interpretation of interactions with the AGQ Assistant, rather than fine-grained evidence of how such assistance shaped moment-by-moment interaction. Future work could incorporate multimodal interaction analysis to better understand how the AGQ Assistant and its suggestions are perceived and negotiated within situated interview dynamics. Second, although our sample included interviewers across proficiency levels, we did not quantitatively compare how novice, intermediate, and advanced interviewers evaluated, adopted, adapted, or rejected AGQs. Future work could recruit a larger and more balanced sample across proficiency levels to examine how interviewer proficiency level shapes the perception and use of AGQs.

\section*{Glossary}
\subsection*{Part 1: Acronyms}
\begin{itemize}
    \item AGQ(s) = AI-generated follow-up question(s)
    \item GPT = generative pre-trained transformer
    \item LLM(s) = large language model(s)
    \item NLP = natural language processing
    \item HCI = human-computer interaction
    \item BERT = bidirectional encoder representations from Transformers
    \item RAG = retrieval-augmented generation
    \item RLHF = reinforcement learning with human feedback
    \item CP(s) = collaboration pattern(s)
\end{itemize}

\subsection*{Part 2: Key Definitions}
\begin{itemize}
    \item \textbf{AI-driven/Reverse Wizard-of-Oz}: An experimental setup where the participant believes they are interacting with a human interviewer, while the questions are actually generated by an AI system and delivered by a human proxy in real-time.
    \item \textbf{Bidirectional Questioning}: A mechanism that allows a two-way information exchange: the AI can prompt the human interviewer for context clarification, and the human can query the AI for specific angles of inquiry.
    \item \textbf{AI Checker}: A safety and quality control module designed to review AI-generated questions for ethical compliance, relevance, and tone before they are presented to the human interviewer or participant.
    \item \textbf{Frontstage AI Model}: An interaction configuration where the AI agent is explicitly presented as an active conversational partner or co-interviewer. In this model, the AI is visible to the participant, directly engaging in the dialogue or having its generated content attributed explicitly to the machine identity.
    \item \textbf{Backstage AI Model}: An interaction configuration where the AI agent operates implicitly behind the scenes, providing real-time support (e.g., generating questions or insights) solely to the human interviewer. In this model, the AI is invisible to the participant, and the human interviewer acts as the sole interface, retaining full agency over information delivery.
    \item \textbf{Theoretical Saturation}: The state reached during the data collection process when incoming participant responses begin to converge on similar narratives, resulting in informational redundancy. It marks the point where further interviews cease to yield novel thematic insights or distinct conceptual categories.
\end{itemize}

\section*{Acknowledgements}
We would like to take this opportunity to express our gratitude to the participants of this study for their valuable insights. We would also like to clarify that we used ChatGPT solely for spell-checking and grammar revisions to enhance content clarity. Our research team assumes full responsibility for the content of this publication. We did not use generative AI for data collection, analysis, or image generation. The study protocol was reviewed and approved by the Institutional Review Board (IRB). Informed consent was obtained from all individual participants involved in the study.


\section*{Data availability}
The data that has been used is confidential.

\section*{Declaration of interests}
The authors declare that they have no known competing financial interests or personal relationships that could have appeared to influence the work reported in this paper.



\appendix






\bibliographystyle{elsarticle-num} 
\bibliography{main,references,references_Iris, revision}


%
%
%
\end{document}